\documentclass[twocolumn,tighten,times,trackchanges]{aastex631}

\newcommand{\editR}[2][black]{{\color{#1} #2}}
\newcommand{\editX}[2][black]{{\color{#1}}}

\usepackage{savesym}
\savesymbol{tablenum}
\usepackage{siunitx}
\restoresymbol{SIX}{tablenum}

\usepackage{tipa}
\usepackage{amsmath}
\usepackage{ulem}

\shorttitle{}
\shortauthors{}
\graphicspath{{./}{figures/}}

\begin{document}

\title{Fine-grained rim formation via kinetic dust aggregation in shock waves around evaporating icy planetesimals}

\author[0000-0003-0947-9962]{Sota Arakawa}
\affiliation{Division of Science, National Astronomical Observatory of Japan, 2-21-1 Osawa, Mitaka, Tokyo, 181-8588, Japan}

\author[0000-0002-1481-6313]{Hiroaki Kaneko}
\affiliation{Department of Earth and Planetary Sciences, Tokyo Institute of Technology, 2-12-1 Ookayama, Meguro, Tokyo, 152-8550, Japan}

\author[0000-0003-3924-6174]{Taishi Nakamoto}
\affiliation{Department of Earth and Planetary Sciences, Tokyo Institute of Technology, 2-12-1 Ookayama, Meguro, Tokyo, 152-8550, Japan}



\begin{abstract}

Fine-grained rims (FGRs) are frequently found around chondrules in primitive chondrites.
The remarkable feature of FGRs is their submicron-sized and non-porous nature.
The typical thickness of FGRs around chondrules is 10--100 \si{\micro}m.
Recently, a novel idea was proposed for the origin of FGRs: high-speed collisions between chondrules and fine dust grains called the kinetic dust aggregation process.
Experimental studies revealed that (sub)micron-sized ceramic particles can stick to a ceramic substrate in a vacuum when the impact velocity is approximately in the range of 0.1--1 km/s.
In this study, we examine the possibility of FGR formation via kinetic dust aggregation in chondrule-forming shock waves.
When shock waves are created by undifferentiated icy planetesimals, fine dust grains would be released from the planetary surface due to evaporation of icy planetesimals.
We consider the dynamics of chondrules behind the shock front and calculate the growth of FGRs via kinetic dust aggregation based on simple one-dimensional calculations.
We found that non-porous FGRs with the thickness of 10--100 \si{\micro}m would be formed in shock waves around evaporating icy planetesimals.

\end{abstract}



\section{introduction}

Fine-grained rims (FGRs) are frequently found around chondrules and calcium-aluminum-rich inclusions \editR{(CAIs)} in primitive chondrites.
FGRs are distinguishable from the interchondrule matrix in optical and scanning electron microscopy images as they have different texture and composition, and the typical thickness of FGRs is on the order of 10--100 \si{\micro}m \citep[e.g.,][]{1984PolRe..35..126M,2018E&PSL.481..201H}.
The physical mechanism that produced these rims is still under debate, and several scenarios have been suggested so far \citep[e.g.,][]{1992GeCoA..56.2873M,2006GeCoA..70.1271T,2012GeCoA..98....1T,2019GeCoA.264..118L}.

The majority of studies assumed that FGRs were formed via the accretion of dust particles onto the surfaces of chondrules/CAIs in the turbulent solar nebula \citep[e.g.,][]{1992GeCoA..56.2873M,1998Icar..134..180M,2004Icar..168..484C,2019Icar..321...99X,2021Icar..35414053X,2021Icar..36714538M,2022Icar..37414726K}.
\editX{The} \editR{This} nebular scenario naturally reproduces the positive correlation between the rim thickness and the chondrule radius, which is reported for FGRs around chondrules in CM chondrites \citep[e.g.,][]{1992GeCoA..56.2873M,2018E&PSL.481..201H,2021GeCoA.295..135Z}.

However, \citet{2019GeCoA.264..118L} pointed out that the nebular scenario has a difficulty explaining the low porosity of FGRs.
Assuming that collisions between chondrules and fine grains occurred in the turbulent solar nebula, the impact velocity would be approximately or lower than 1 m/s and porous dust rims with the porosity of approximately 60\% would be formed \citep{2013GeCoA.116...41B}.
In addition, \editX{(sub)micron-sized} dust grains turned into fluffy aggregates prior to the accretion onto chondrules when the grain size is smaller than 1 \si{\micro}m \citep[e.g.,][]{2017ApJ...846..118A,2019ApJ...887..248M,2022Icar..37414726K}.
The typical grain size of FGRs in primitive chondrites is indeed submicron \citep[e.g.,][]{2000GeCoA..64.3263L,2008GeCoA..72..602C,2021GeCoA.295..135Z}, although grain size might be subsequently modified by aqueous/thermal alteration processes.
Hence the structure of FGRs formed in the turbulent solar nebula would be highly porous; which seems to be inconsistent with the observed compact FGRs with low porosity of 10--20\% \citep[e.g.,][]{2006GeCoA..70.1271T}.

Alternatively, several studies investigated a scenario that FGRs were formed after accretion of chondrite parent bodies \citep[e.g.,][]{1993Metic..28..669S,2006GeCoA..70.1271T,2010GeCoA..74.4438T,2012GeCoA..98....1T}.
In the framework of \editX{the} \editR{this} parent-body scenario, the FGRs are formed via aqueous/thermal alterations of host chondrules and/or via impact-induced compaction/fragmentation of the matrix material around chondrules \citep[see][and references therein]{2010GeCoA..74.4438T}.
The parent-body scenario can naturally explain the non-porous nature of FGRs, and this is one of the reasons why parent-body scenario is still favored for the origin of FGRs.
However, another difficulty exists when we consider the parent-body scenario.
Based on the fabric analysis by high-resolution electron backscatter diffraction, \citet{2011NatGe...4..244B} found that FGRs were exposed to a spherically symmetric stress field while the matrix exhibits a bulk uniaxial stress field.
This result indicates that FGRs were compressed prior to rimmed chondrules being incorporated into chondrite parent bodies.
Moreover, \citet{2013Icar..225..558B} revealed that impact-induced compaction cannot form non-porous FGRs, based on their impact experiments into mixtures of chondrule analogs and fine dust particles.

\begin{figure}
\centering
\includegraphics[width=\columnwidth]{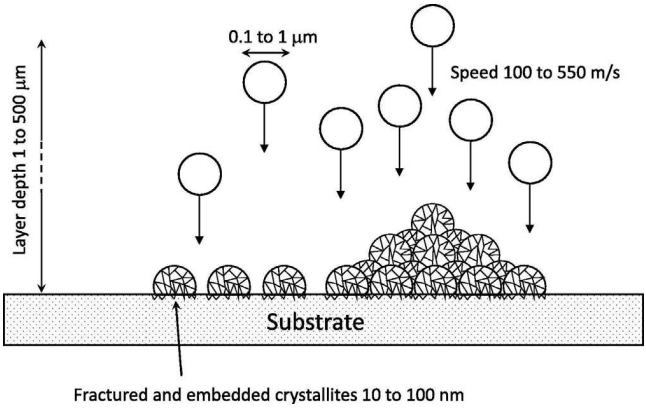}
\caption{
Illustration of the fracturing and compaction of dust particles during kinetic dust aggregation.
Note that the maximum/minimum velocities for adhesion shown in this figure (100 and 550 m/s) are for the case of (sub)micron-sized yttrium iron garnet (${\rm Y}_{3}{\rm Fe}_{5}{\rm O}_{12}$) particles, and these critical velocities should depend on the composition and grain size in reality.
Figure taken from \citet{2019GeCoA.264..118L} modified after \citet{2014APExp...7c5501J}.
}
\label{fig:Liffman}
\end{figure}

To solve these problems, \citet{2019GeCoA.264..118L} proposed a novel idea for the origin of FGRs: high-speed collisions between chondrules and fine dust grains called the {\it kinetic dust aggregation} process.
The kinetic dust aggregation is also known as the aerosol deposition method \citep[e.g.,][]{akedo2006aerosol,akedo2008room,akedo2008aerosol,2014APExp...7c5501J,hanft2015overview} in the field of ceramic coating technologies.
Experimental studies revealed that (sub)micron-sized ceramic particles can stick to a ceramic substrate in a vacuum, and the impact velocity for sticking is approximately 0.1--1 km/s \citep[see][and references therein]{hanft2015overview}.
Molecular dynamics simulations also confirmed that 10--100 nm-sized brittle nanoparticles can stick to the substrate when the impact velocity is on the order of 0.1--1 km/s \citep[e.g.,][]{2014JTST...23..541D}.
The resulting dust layer formed via the kinetic dust aggregation have low porosity and are fine grained, as illustrated in Figure \ref{fig:Liffman}.
Therefore, we can reproduce the observed structure of FGRs if they are formed via the kinetic dust aggregation process, which should be related to chondrule-forming supersonic events.

In this study, we examine the possibility of FGR formation via kinetic dust aggregation in chondrule-forming shock waves.
Shock waves caused by eccentric planetesimals in the gaseous solar nebula is one of the leading candidates for the chondrule-forming transient events \citep[e.g.,][]{1998Sci...279..681W,2004M&PS...39.1809C,2012ApJ...752...27M,2016ApJ...818..103M,2018ApJ...857...96M,2019ApJ...871..110N}.
When shock waves are created by undifferentiated icy planetesimals, fine dust grains would be released from the planetary surface due to evaporation of icy planetesimals \citep[e.g.,][]{2013ApJ...764..120T}.
The enrichment of fine dust grains in chondrule-forming environment would be preferred from a variety of perspectives \citep[e.g.,][]{2008Sci...320.1617A,2012GeCoA..78....1H,2015GeCoA.148..228T}.
Based on the oxygen isotope composition and oxidation state of chondrule olivine, \citet{2013GeCoA.101..302S} concluded that chondrules in CR chondrites formed under ${\rm H}_{2}{\rm O} / {\rm H}_{2}$ ratios between 10 and 1000 times the solar ratio \citep[see also][]{2015GeCoA.148..228T}.
As evaporating icy planetesimals can supply high ${\rm H}_{2}{\rm O}$ vapor pressure, our scenario is also consistent with the observed oxygen fugacity.
We consider the dynamics of chondrules behind the shock front and calculate the growth of FGRs via kinetic dust aggregation.
Although our numerical results are based on simple one-dimensional calculations, we found that non-porous FGRs with the thickness of 10--100 \si{\micro}m would be formed in shock waves around evaporating icy planetesimals.

\section{model}

\subsection{Outline}

\begin{figure}
\centering
\includegraphics[width=\columnwidth]{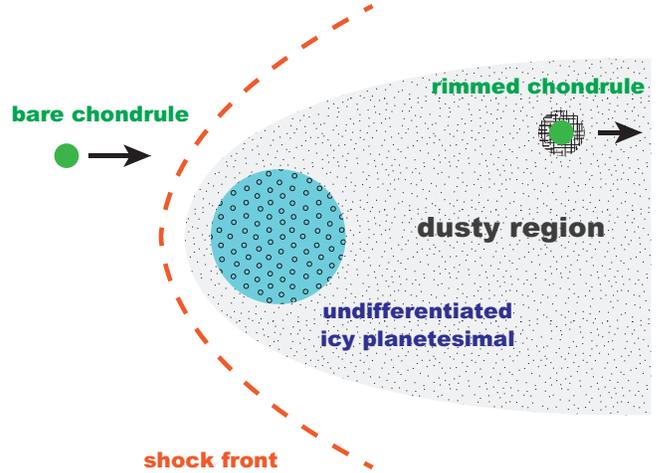}
\caption{
Schematic of our fine-grained rim formation scenario.
Evaporation of undifferentiated icy planetesimals produces dusty regions behind the shock front.
As chondrules entered the shock wave have a relative velocity with respect to fine grains, which are dynamically coupled with gas, fine dust grains collide with chondrules and fine-grained rim will be formed in dusty regions.
}
\label{fig:schematic}
\end{figure}

The formation process of FGRs in shock waves is illustrated in Figure \ref{fig:schematic}.
We consider the accretion of FGRs onto bare chondrules.
When shock waves are caused by undifferentiated icy planetesimals, the dusty region would be formed behind the shock front due to evaporation of planetesimals.
We assume that fine dust grains released from planetesimals are dynamically coupled with gas while chondrules entered the shock wave have relative velocity with respect to gas, and fine dust grains collide with chondrules.
Then fine dust grains accrete onto chondrules if the impact velocity satisfies the condition for adhesion.

We briefly explain the models and settings in the following sections.
In this study, we discuss the dynamics of chondrules in one-dimensional normal shocks.
The basic framework of our model is identical to that used in \citet{2019ApJ...877...84A}.
We calculate the evolution of the velocity and radius of rimmed chondrules, $v$ and $r$, simultaneously.

\subsection{Gas structure}

We do not calculate the dynamics of gas behind the shock front but assume a simple gas structure.
Then the dynamics of chondrules is simulated in the given gas flow.
We assume that the gas velocity with respect to the shock front, $v_{\rm g}$, and the gas density, $\rho_{\rm g}$, evolve as functions of the distance from the shock front, $x$:
\begin{equation}
v_{\rm g} = 
\begin{cases}
\displaystyle v_{0} & {( x < 0 )}, \\
\displaystyle v_{0} + {\left( v_{\rm post} - v_{0} \right)} \exp{\left( {- x}/{L} \right)} & {( x \geq 0 )},
\end{cases}
\label{eq:vg}
\end{equation}
and
\begin{equation}
\rho_{\rm g} = \frac{v_{0}}{v_{\rm g}} \rho_{{\rm g}, 0},
\end{equation}
where $v_{0}$ is the pre-shock gas velocity with respect to the shock front, $v_{\rm post}$ is the post-shock gas velocity with respect to the shock front, $\rho_{{\rm g}, 0}$ is the pre-shock gas density, and $L$ is the spatial scale of the shock.
The spatial scale of the shock should be several times or much larger than the radius of planetesimals, $r_{\rm p}$ \citep[see][and references therein]{2019ApJ...877...84A}.
However, the value of $L$ should also depend on the physical properties of the solar nebula, e.g., the turbulence strength and the opacity.
Thus we regard $L$ as a parameter and consider three cases: $L = 3 \times 10^{4}\ {\rm km}$, $1 \times 10^{4}\ {\rm km}$, and $3 \times 10^{3}\ {\rm km}$.
The post-shock gas velocity, $v_{\rm post}$, is given by $v_{\rm post} = {\left[ {(\gamma - 1)}/{(\gamma + 1)} \right]} v_{0}$, where $\gamma$ is the ratio of specific heats.
We set $\rho_{{\rm g}, 0} = 5 \times 10^{-10}\ {\rm g}\ {\rm cm}^{-3}$, $v_{0} = 12\ {\rm km}\ {\rm s}^{-1}$, and $\gamma = 1.4$.
Similarly, the temperature of the gas $T_{\rm g}$ is assumed as follows:
\begin{equation}
T_{\rm g} = 
\begin{cases}
\displaystyle T_{0} & {( x < 0 )}, \\
\displaystyle T_{0} + {\left( T_{\rm post} - T_{0} \right)} \exp{\left( {- x}/{L} \right)} & {( x \geq 0 )}.
\end{cases}
\end{equation}
We assume that the pre-shock gas temperature is $T_{0} = 200\ {\rm K}$ and the post-shock gas temperature is $T_{\rm post} = 1600\ {\rm K}$.
The most probable molecular velocity $c_{\rm s}$ is given by $c_{\rm s} \equiv {(2 k_{\rm B} T_{\rm g} / m_{\rm g})}^{1/2} = 1.3\ {\left[ T_{\rm g} / {( 200\ {\rm K} )} \right]}^{1/2}\ {\rm km}\ {\rm s}^{-1}$, where $k_{\rm B} = 1.38 \times 10^{-16}\ {\rm erg}\ {\rm K}^{-1}$ is the Boltzmann constant and $m_{\rm g} = 3.34 \times 10^{-24}\ {\rm g}$ is the gas molecule mass, which value corresponds to ${\rm H}_{2}$ gas.

\subsection{Chondrule dynamics}

The velocity of chondrules with respect to the shock front, $v$, will change as follows \citep[e.g.,][]{1991Icar...93..259H}:
\begin{equation}
\frac{4 \pi}{3} r^{3} \rho \frac{{\rm d}v}{{\rm d}x} = - \frac{C_{\rm D}}{2} \pi r^{2} \rho_{\rm g} \frac{\left| v - v_{\rm g} \right|}{v} {\left( v - v_{\rm g} \right)},
\end{equation}
where $C_{\rm D}$ is the drag coefficient, $r$ is the chondrule radius, and $\rho = 3.3\ {\rm g}\ {\rm cm}^{-3}$ is the internal density of chondrules \citep{2004M&PS...39.1809C}.
Assuming that the temperature of chondrules is equal to gas temperature, the drag coefficient, $C_{\rm D}$, is given by
{\footnotesize
\begin{equation}
C_{\rm D} = \frac{2 \sqrt{\pi}}{3 s} + \frac{2 s^{2} + 1}{\sqrt{\pi} s^{3}} \exp{(- s^{2})} + \frac{4 s^{4} + 4 s^{2} - 1}{2 s^{4}} {\rm erf}{(s)},
\end{equation}
}where the Mach number, $s$, is given by $s \equiv {|v - v_{\rm g}|} / c_{\rm s}$.
Here we introduce the stopping length of chondrules, $l_{\rm stop}$.
For the case in which chondrules move in gas with supersonic velocities, $l_{\rm stop}$ is approximately given by
\begin{eqnarray}
l_{\rm stop} &\equiv& {\left( \frac{1}{v} {\left| \frac{{\rm d}v}{{\rm d}x} \right|} \right)}^{-1} \nonumber \\
             &\simeq& \frac{4}{3} \frac{\rho}{\rho_{\rm g}} {\left( \frac{v - v_{\rm g}}{v} \right)}^{-2} r.
\label{eq:lstop}
\end{eqnarray}
If the spatial scale of shock is much larger than the stopping length ($L \gg l_{\rm stop}$), the velocity of a chondrule reaches $v \simeq v_{\rm post}$ behind the shock front, while $v$ barely changes when $L \ll l_{\rm stop}$ \citep[see][]{2019ApJ...877...84A}.
On the other hand, for the case in which chondrules move in gas with subsonic velocities, $l_{\rm stop}$ is approximately given by the following equation:
\begin{equation}
l_{\rm stop} \simeq 0.64 \frac{\rho}{\rho_{\rm g}} {\left( \frac{c_{\rm s} {\left| v - v_{\rm g} \right|}}{v^{2}} \right)}^{-1} r.
\label{eq:lstop2}
\end{equation}

\subsection{Accretion of fine-grained rims}

In this study, we calculate the accretion of fine-grained rim in shock waves. 
The mass accretion rate per unit length, ${{\rm d}m}/{{\rm d}x}$, is given by
\begin{equation}
\frac{{\rm d}m}{{\rm d}x} = Q \rho_{\rm d} \pi r^{2} \frac{v_{\rm imp}}{v},
\end{equation}
where $Q$ is the coefficient for adhesion/erosion of fine grains, and $\rho_{\rm d}$ is the dust density.
Here we assume that fine grains are both dynamically and thermally coupled with gas, and the impact velocity of fine grains is given by
\begin{equation}
v_{\rm imp} = {\left| v - v_{\rm g} \right|}.
\end{equation}
The growth rate of the thickness of rims, ${{\rm d}r}/{{\rm d}x}$, is given by the following equation:
\begin{equation}
\frac{{\rm d}r}{{\rm d}x} = \frac{1}{4 \pi \rho r^{2}} \frac{{\rm d}m}{{\rm d}x},
\end{equation}
and we do not consider the porosity of FGRs for simplicity.\footnote{
\editR{The porosity of FGRs formed via the kinetic dust aggregation process would be 10\% or less \citep[e.g.,][]{hanft2015overview}, although it must depend on many parameters including the impact velosity and  the material composition.}
}
The thickness of the rim, $\Delta$, is given by
\begin{equation}
\Delta = r - r_{0},
\end{equation}
where $r_{0}$ is the radius of the bare chondrule.

The coefficient for adhesion/erosion depends on the impact velocity: $Q = Q {\left( v_{\rm imp} \right)}$.
In this study, we assume that $Q {\left( v_{\rm imp} \right)}$ is given by a step function as follows:
\begin{equation}
Q = 
\begin{cases}
\displaystyle Q_{\rm ad} & {\left( v_{\rm min} \le v_{\rm imp} \le v_{\rm max} \right)}, \\
\displaystyle Q_{\rm er} & {\left( v_{\rm imp} > v_{\rm max}\ {\rm and}\ \Delta > 0 \right)}, \\
0 & {\left( {\rm otherwise} \right)},
\end{cases}
\end{equation}
where $Q_{\rm ad}$ and $Q_{\rm er}$ are the coefficients for adhesion/erosion, and $v_{\rm max}$ and $v_{\rm min}$ are the maximum/minimum velocity for adhesion, respectively.
We change the values of $Q_{\rm ad}$, $Q_{\rm er}$, $v_{\rm max}$, and $v_{\rm min}$ as parameters (see Table \ref{table:coeff}).

\begin{table*}
\caption{
Fundamental parameters for describing the accretion of FGRs: $Q_{\rm ad}$, $Q_{\rm er}$, $v_{\rm max}$, and $v_{\rm min}$.
}
\label{table:coeff}
\centering
\begin{tabular}{ccc}
{\bf Parameter} & {\bf Symbol} & {\bf Value} \\ \hline
Coefficient for adhesion & $Q_{\rm ad}$ & $0.5$ or $0.2$ \\
Coefficient for erosion & $Q_{\rm er}$ & $0$ or $-1$ \\
Maximum velocity for adhesion & $v_{\rm max}$ & $1\ {\rm km}\ {\rm s}^{-1}$ or $0.3\ {\rm km}\ {\rm s}^{-1}$ \\
Minimum velocity for adhesion & $v_{\rm min}$ & $0.1\ {\rm km}\ {\rm s}^{-1}$ or $0.3\ {\rm km}\ {\rm s}^{-1}$
\end{tabular}
\end{table*}

We do not consider the erosion of chondrules for simplicity; however, it might play an important role for the origin of a non-zero constant in the linear relationship between $\Delta$ and $r_{0}$ reported from observations of chondrules in CM chondrites \citep{2019GeCoA.264..118L}.
The erosion of chondrules may also be problematic in the context of the survival of chondrules in shock waves if $Q_{\rm er} \ll -1$ \citep[e.g.,][]{2014ApJ...797...30J}.
However, we can imagine that the value of $Q_{\rm er}$ for the erosion of chondrules should differ from that for the erosion of FGRs, and our knowledge of erosion of chondrules is still limited.
Thus, future studies on the physics of erosive collision are necessary.

\subsection{Production of silicate dust from evaporating planetesimals}

We simply set the following assumption for the structure of $\rho_{\rm d}$:
\begin{equation}
\rho_{\rm d} = 
\begin{cases}
0 & {\left( x < 0 \right)}, \\
\displaystyle \chi \rho_{\rm g} & {\left( x \geq 0 \right)},
\end{cases}
\end{equation}
where $\chi$ is the dust-to-gas mass ratio in the dusty region formed behind the shock front.
In this study, we set $\chi = 1$ based on the order-of-magnitude analysis shown below.

In this study, we consider the evaporation of undifferentiated icy planetesimals.
The planetesimal surface is heated by a hot shocked gas, and the surface ice evaporates.
For the case of the supersonic limit, \citet{2013ApJ...764..120T} derived that the evaporation flux of the surface ice of the planetesimal is approximately given by
\begin{equation}
J_{\rm ice} \simeq \pi {R_{\rm p}}^{2} \frac{2 \gamma}{{\left( \gamma + 1 \right)}^{2}} \frac{\alpha \rho_{{\rm g}, 0} {v_{0}}^{3}}{L_{\rm eva}},
\end{equation}
where $L_{\rm eva} = 2.7 \times 10^{10}\ {\rm erg}\ {\rm g}^{-1}$ is the latent heat of evaporation of ice, and $\alpha$ is the non-dimensional parameter called the Stanton number, which expresses the efficiency of heat conduction. 
\citet{2013ApJ...764..120T} found that the realistic range of $\alpha$ for planetesimal bow shocks is $10^{-2} \le \alpha \le 10^{-1}$.
When the surface ice evaporates, dust grains are also released from the surface of undifferentiated planetesimals.
The mass flux of the released dust grains, $J_{\rm dust}$, would be simply given as follows:
\begin{equation}
J_{\rm dust} = f_{\rm dust/ice} J_{\rm ice},
\end{equation}
where $f_{\rm dust/ice}$ is the dust-to-ice mass ratio of the evaporating undifferentiated planetesimals.

The value of $f_{\rm dust/ice}$ is uncertain; however, several studies on the internal structure of comet 67P/Churyumov--Gerasimenko suggested that the dust-to-ice mass ratio of the comet is significantly higher than one, $f_{\rm dust/ice} \gg 1$ \citep[e.g., ][]{2019MNRAS.482.3326F,2019MNRAS.483.2337P,2020MNRAS.497.1166A}.
The bulk density of the comet indicates $f_{\rm dust/ice} \simeq 9$ \citep{2020MNRAS.497.1166A} if comets are formed via gravitational collapse of a cloud of dust aggregates in the solar nebula \citep[e.g.,][]{2012Icar..221....1S,2017MNRAS.469S.149W,2021A&A...647A.126V}.
\citet{2019MNRAS.482.3326F} also reviewed the dust-to-ice mass ratio of other comet nuclei visited by space missions and of trans-Neptunian objects (TNOs), and these objects have generally the value of $f_{\rm dust/ice} \gg 3$.

These estimates on the value of $f_{\rm dust/ice}$ are an order of magnitude higher than the classical value for the dust composition in protoplanetary disks \citep[e.g.,][]{1994ApJ...421..615P,2001ApJ...553..321D}.
We note, however, that recent studies on the dust composition of protoplanetary disks \citep[see][and references therein]{2018ApJ...869L..45B} suggest that $f_{\rm dust/ice}$ should be several times higher than that predicted by \citet{1994ApJ...421..615P}.
\citet{2021ApJ...910...26T} also evaluated the dust-to-ice mass ratio using the scattering polarization in the envelope of the low mass protostar L1551 IRS 5, and they found that icy dust grains with the radius of a few \si{\micro}m (or larger) and $f_{\rm dust/ice} \gtrsim 10$ are consistent with the observed polarization excess around a wavelength of 3 \si{\micro}m.
Thus, we can expect that icy planetesimals are formed from dust-rich icy grains with $f_{\rm dust/ice} \gg 1$.

Assuming the mass conservation, the dust density is given by
\begin{equation}
\rho_{\rm d} \simeq \frac{J_{\rm dust}}{\pi {R_{\rm d}}^{2} v_{\rm g}},
\end{equation}
where $R_{\rm d}$ is the radius of the dusty region.
Then, the typical value of the dust-to-gas mass ratio behind the shock front would be obtained as follows:
{\footnotesize
\begin{eqnarray}
\chi & \simeq & f_{\rm dust/ice} {\left( \frac{R_{\rm p}}{R_{\rm d}} \right)}^{2} \frac{2 \gamma}{{\left( \gamma + 1 \right)}^{2}} \frac{\alpha {v_{0}}^{2}}{L_{\rm eva}} \nonumber \\
     & \simeq & 0.8 {\left( \frac{f_{\rm dust/ice}}{9} \right)} {\left( \frac{R_{\rm d} / R_{\rm p}}{3} \right)}^{-2} {\left( \frac{\alpha}{0.03} \right)} {\left( \frac{v_{0}}{12\ {\rm km}\ {\rm s}^{-1}} \right)}^{2}.
\label{eq:chi}
\end{eqnarray}
}Therefore, the value of $\chi \simeq 1$ could be achieved in the dusty region caused by the evaporation of undifferentiated icy planetesimals, although there are large uncertainties of the values of $f_{\rm dust/ice}$, $R_{\rm p} / R_{\rm d}$, and $\alpha$.
Thus, future studies on the detailed analysis on the dust-to-gas mass ratio behind the shock front would be essential.

The diameter-density relation among TNOs are investigated so far \citep[e.g.,][]{2012AREPS..40..467B,2019Icar..334...30G}.
Large TNOs whose diameter is larger than 1000 km have usually the bulk density of approximately $2$--$3\ {\rm g}\ {\rm cm}^{-3}$, while mid-sized TNOs with a diameter smaller than 1000 km have the bulk density of approximately $1\ {\rm g}\ {\rm cm}^{-3}$.
\citet{2019Icar..334...30G} pointed out that difference in bulk density may reflect the porosity change.
Thus, icy planetesimals with a diameter smaller than 1000 km would be porous and undifferentiated bodies, and the dusty region may be formed when shock waves are caused by these mid-sized planetesimals.
In contrast, large icy bodies with a diameter larger than 1000 km would be differentiated and might not be suitable for the formation of rimmed chondrules.

\section{results}

\subsection{Impact velocity}

\begin{figure}
\centering
\includegraphics[width=\columnwidth]{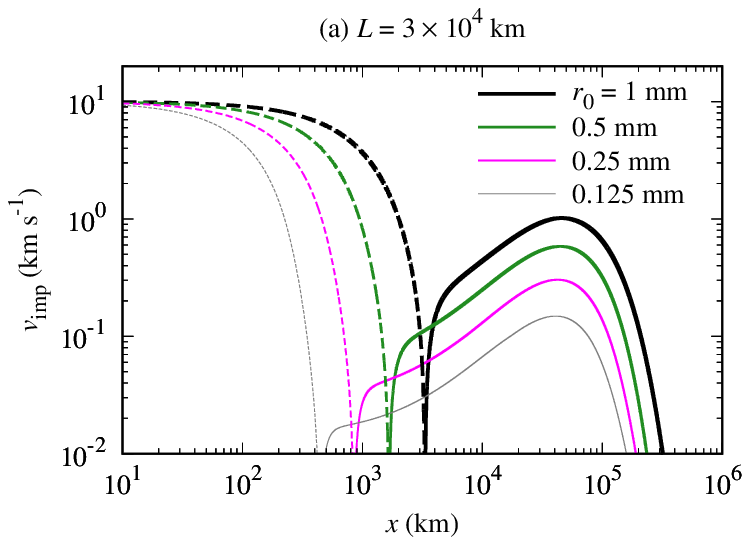}
\includegraphics[width=\columnwidth]{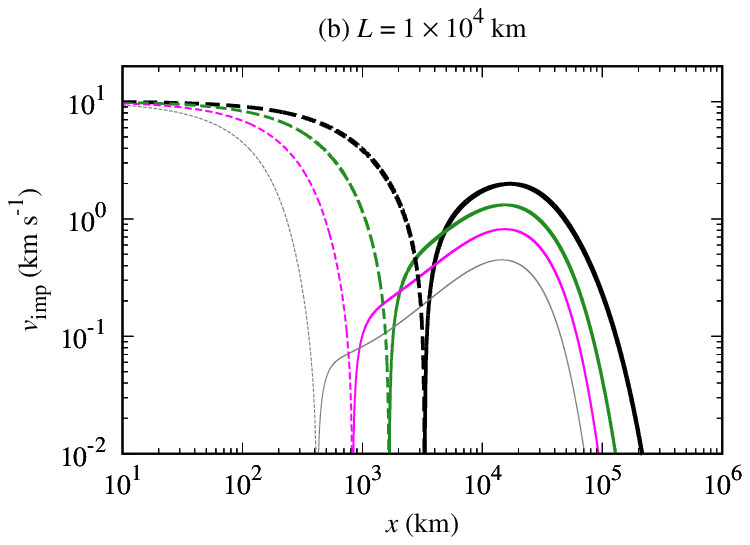}
\includegraphics[width=\columnwidth]{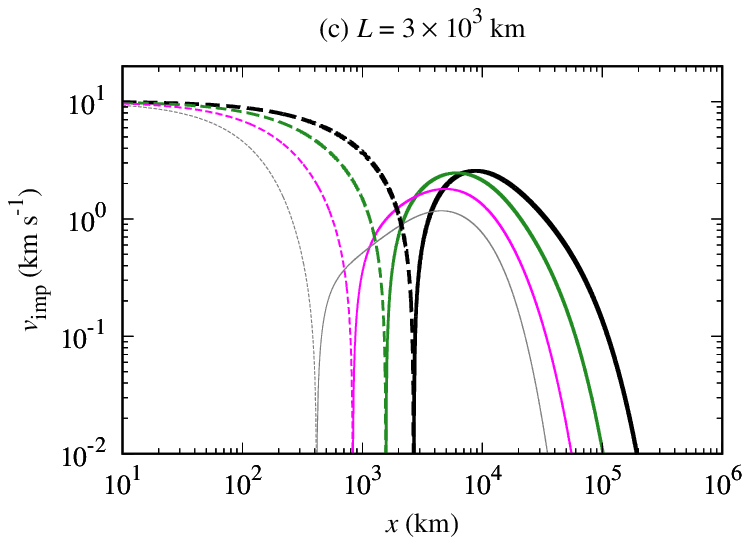}
\caption{
Impact velocity of fine grains, $v_{\rm imp} = {\left| v - v_{\rm g} \right|}$.
(a) For the case of $L = 3 \times 10^{4}\ {\rm km}$.
(b) For the case of $L = 1 \times 10^{4}\ {\rm km}$.
(c) For the case of $L = 3 \times 10^{3}\ {\rm km}$.
Solid lines indicate $v - v_{\rm g} < 0$ while dashed lines indicate $v - v_{\rm g} > 0$.
We set $Q_{\rm ad} = 0.5$, $Q_{\rm er} = 0$, $v_{\rm max} = 1\ {\rm km}\ {\rm s}^{-1}$, and $v_{\rm min} = 0.1\ {\rm km}\ {\rm s}^{-1}$.
}
\label{fig:vimp}
\end{figure}

First, we show the impact velocity of fine grains.
Figure \ref{fig:vimp} shows $v_{\rm imp}$ as a function of the distance from the shock front.
Panels (a), (b), and (c) show the results for the cases of $L = 3 \times 10^{4}\ {\rm km}$, $L = 1 \times 10^{4}\ {\rm km}$, and $L = 3 \times 10^{3}\ {\rm km}$, respectively.
Solid lines indicate $v - v_{\rm g} < 0$ while dashed line indicate $v - v_{\rm g} > 0$.

\citet{2019ApJ...877...84A} found that the dynamical evolution of chondrules in shock waves can be divided into two stages: deceleration region behind the shock front (Stage 1) and recovery region where the velocity of chondrules and gas approach the pre-shock velocity (Stage 2).
As shown in Figure \ref{fig:vimp}, the change of Stages 1/2 occurred at around $x \sim 1000\ {\rm km}$ for the case of $\rho_{{\rm g}, 0} = 5 \times 10^{-10}\ {\rm g}\ {\rm cm}^{-3}$, and small chondrules enter Stage 2 earlier than larger chondrules.
This is because smaller chondrules have shorter stopping lengths (see Equations \ref{eq:lstop} and \ref{eq:lstop2}).
For the cases of $L \ge 1 \times 10^{4}\ {\rm km}$, $v_{\rm imp}$ in Stage 2 is approximately proportional to the radius of the bare chondrule $r_{0}$.
In Discussion section, we will derive $v_{\rm imp} = v_{\rm imp} {\left( r_{0} \right)}$ in Stage 2 from an analytical argument.

\subsection{Evolution of rim thickness}

Then, we show the evolution of the thickness of FGRs in the dusty region.
We introduce the results for two cases: rim formation without erosion ($Q_{\rm er} = 0$) and with erosion  ($Q_{\rm er} = -1$).

\subsubsection{Rim formation without erosion}

\begin{figure}
\centering
\includegraphics[width=\columnwidth]{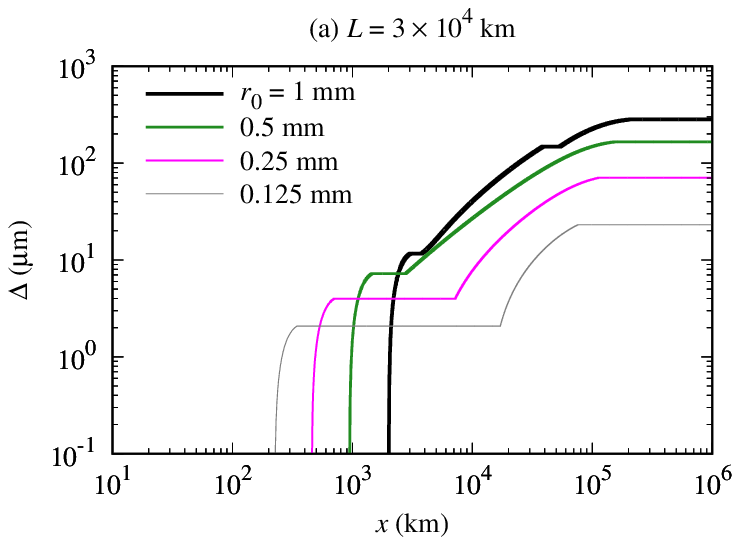}
\includegraphics[width=\columnwidth]{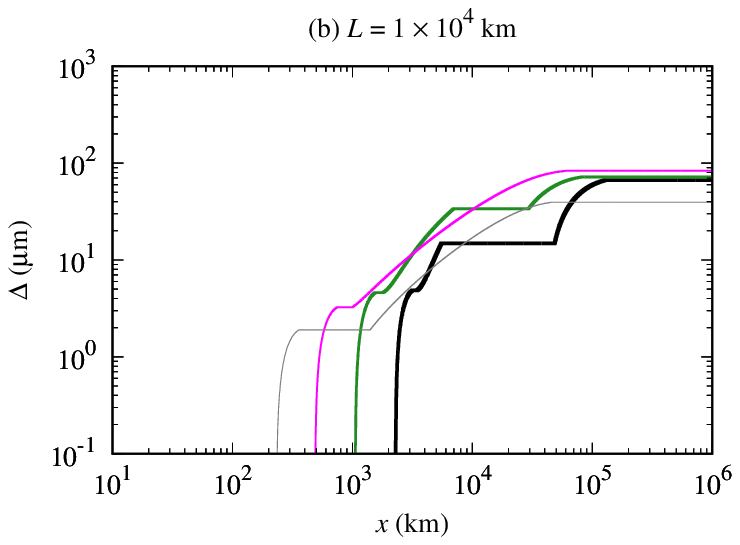}
\includegraphics[width=\columnwidth]{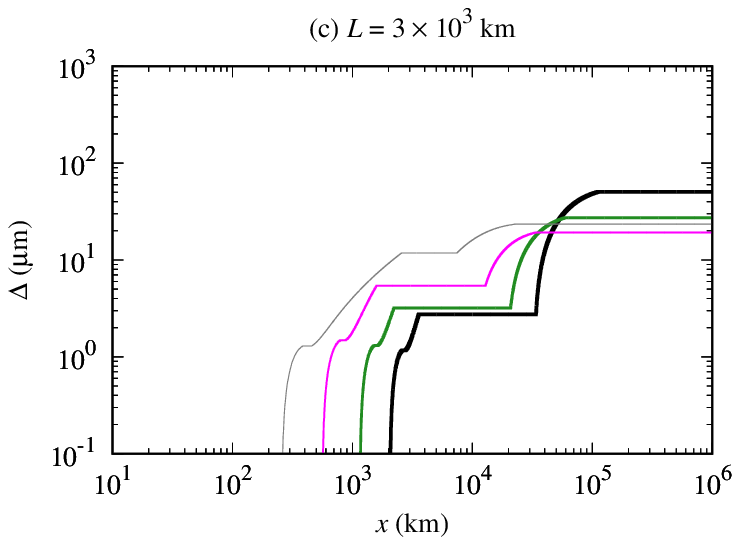}
\caption{
Thickness of fine-grained rims, $\Delta - r - r_{0}$.
(a) For the case of $L = 3 \times 10^{4}\ {\rm km}$.
(b) For the case of $L = 1 \times 10^{4}\ {\rm km}$.
(c) For the case of $L = 3 \times 10^{3}\ {\rm km}$.
We set $Q_{\rm ad} = 0.5$, $Q_{\rm er} = 0$, $v_{\rm max} = 1\ {\rm km}\ {\rm s}^{-1}$, and $v_{\rm min} = 0.1\ {\rm km}\ {\rm s}^{-1}$, and rim formation without erosion is assumed.
}
\label{fig:Delta-x-no-erosion}
\end{figure}

Figure \ref{fig:Delta-x-no-erosion} shows the thickness of FGRs, $\Delta$, as a function of $x$ and $r_{0}$.
Panels (a), (b), and (c) show the results for the cases of $L = 3 \times 10^{4}\ {\rm km}$, $L = 1 \times 10^{4}\ {\rm km}$, and $L = 3 \times 10^{3}\ {\rm km}$, respectively.
Here we set $Q_{\rm ad} = 0.5$, $Q_{\rm er} = 0$, $v_{\rm max} = 1\ {\rm km}\ {\rm s}^{-1}$, and $v_{\rm min} = 0.1\ {\rm km}\ {\rm s}^{-1}$.

As shown in Figure \ref{fig:Delta-x-no-erosion}, FGRs with thickness of 10--100 \si{\micro}m are formed via the kinetic dust aggregation process.
We found that the thickness of FGRs formed in Stage 1 is significantly smaller than the final thickness in these simulations; therefore the FGRs are mainly formed in Stage 2.
In addition, for the case of large $L = 3 \times 10^{4}\ {\rm km}$, the thickness is approximately proportional to $r_{0}$.
We derived analytical solutions for the rim thickness formed in Stages 1 and 2 in Discussion section, and the analytical solutions reproduce the linear relationship between $\Delta$ and $r_{0}$.

\subsubsection{Rim formation with erosion}

\begin{figure}
\centering
\includegraphics[width=\columnwidth]{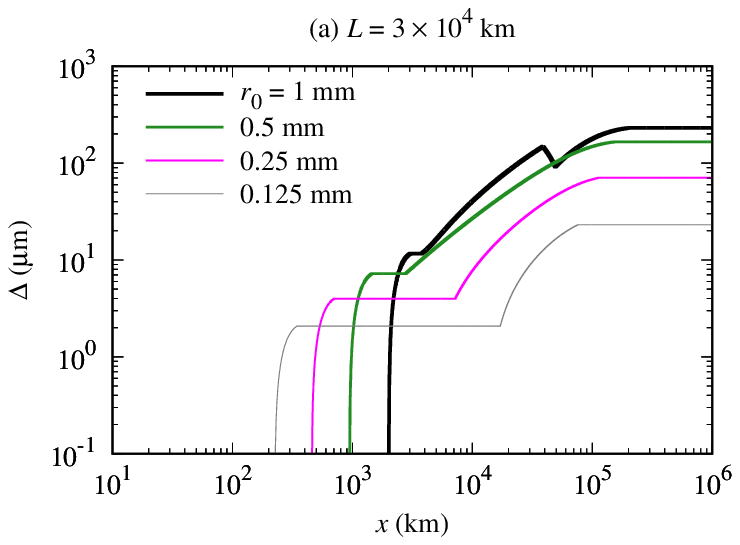}
\includegraphics[width=\columnwidth]{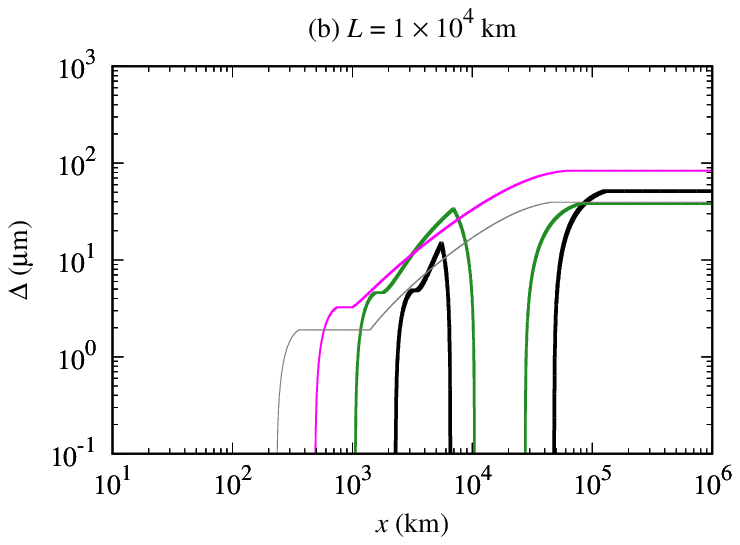}
\includegraphics[width=\columnwidth]{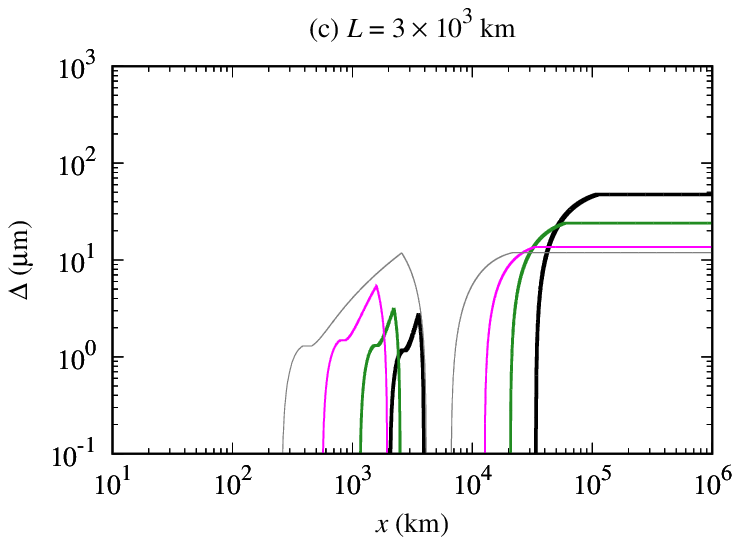}
\caption{
Thickness of fine-grained rims, $\Delta - r - r_{0}$.
(a) For the case of $L = 3 \times 10^{4}\ {\rm km}$.
(b) For the case of $L = 1 \times 10^{4}\ {\rm km}$.
(c) For the case of $L = 3 \times 10^{3}\ {\rm km}$.
We set $Q_{\rm ad} = 0.5$, $Q_{\rm er} = -1$, $v_{\rm max} = 1\ {\rm km}\ {\rm s}^{-1}$, and $v_{\rm min} = 0.1\ {\rm km}\ {\rm s}^{-1}$, and rim formation with erosion is assumed.
}
\label{fig:Delta-x-erosion}
\end{figure}

However, in reality, FGRs would be eroded when $v_{\rm imp}$ is higher than the critical value for erosion.
Although the exact value of the coefficient for erosion, $Q_{\rm er}$, is highly uncertain, the assumption of $Q_{\rm er} < 0$ seems to be more realistic than $Q_{\rm er} = 0$.
Figure \ref{fig:Delta-x-erosion} shows the thickness of FGRs, $\Delta$, as a function of $x$ and $r_{0}$.
Panels (a), (b), and (c) show the results for the cases of $L = 3 \times 10^{4}\ {\rm km}$, $L = 1 \times 10^{4}\ {\rm km}$, and $L = 3 \times 10^{3}\ {\rm km}$, respectively.
We set $Q_{\rm ad} = 0.5$, $Q_{\rm er} = -1$, $v_{\rm max} = 1\ {\rm km}\ {\rm s}^{-1}$, and $v_{\rm min} = 0.1\ {\rm km}\ {\rm s}^{-1}$.

Figure \ref{fig:Delta-x-erosion}(a) shows the evolution of $\Delta$ for the case of $L = 3 \times 10^{4}\ {\rm km}$.
For the case of $r_{0} = 1\ {\rm mm}$ (black line), the erosion of FGRs occurs at around $x \simeq 5 \times 10^{4}\ {\rm km}$ but FGRs partly survive after erosion.
Then fine dust grains accrete onto chondrules again; multi-layered FGRs would be formed by single shock-heating event.
Interestingly, many chondrules in Kivesvaara CM2 chondrite are covered by multi-layered FGRs \citep{1992GeCoA..56.2873M} and our scenario might explain the origin of these multi-layered FGRs.
Our scenario also indicates that inner rims formed in a hotter environment than outer rims.
This would be consistent with the observed characteristics of inner rims \citep[e.g., silicate sintering, sulfides growth, and compaction;][]{2021GeCoA.295..135Z}.

Figure \ref{fig:Delta-x-erosion}(b) shows the evolution of $\Delta$ for the case of $L = 1 \times 10^{4}\ {\rm km}$.
For the cases of $r_{0} = 1\ {\rm mm}$ (black line) and $r_{0} = 0.5\ {\rm mm}$  (green line), FGRs formed before erosion are completely eroded once, then re-accretion of FGRs occurs.
Similar evolutionary path are also found in Figure \ref{fig:Delta-x-erosion}(c), i.e., for the case of $L = 3 \times 10^{3}\ {\rm km}$.
We note that the final thickness of FGRs is in the range of 10--100 \si{\micro}m even if we take into account the effect of erosion.
This is because the final thickness of FGRs is mainly controlled by the accretion of fine grains in Stage 2.

\subsection{Dependence of final rim thickness on chondrule radius}

\begin{figure*}
\centering
\includegraphics[width=\columnwidth]{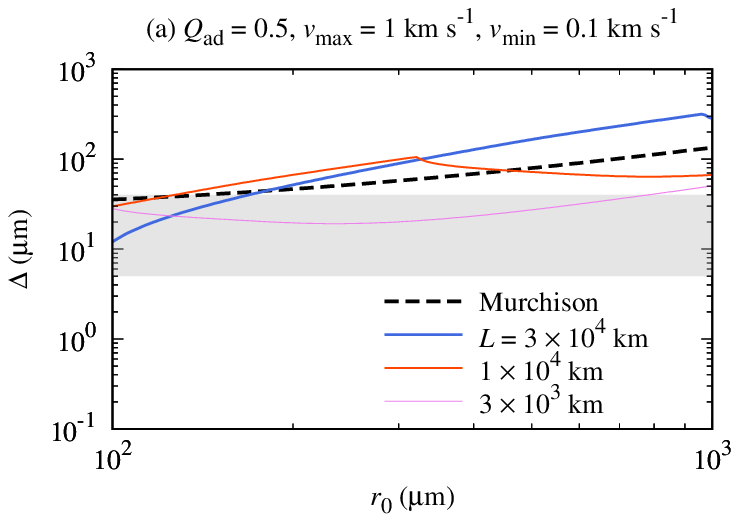}
\includegraphics[width=\columnwidth]{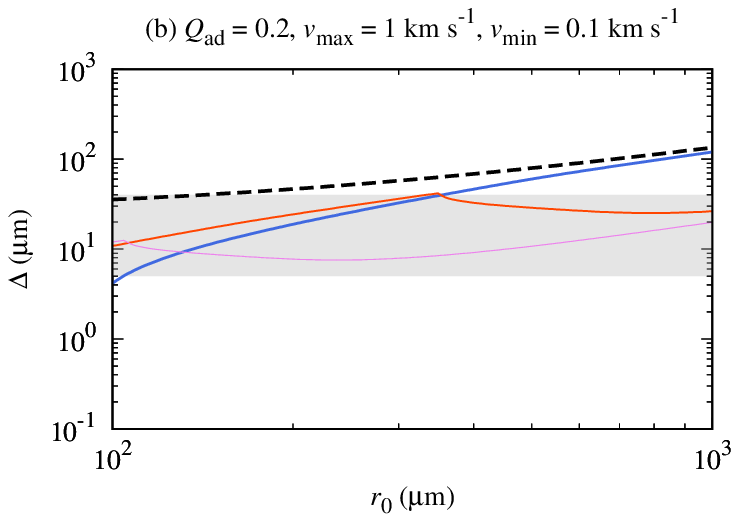}
\includegraphics[width=\columnwidth]{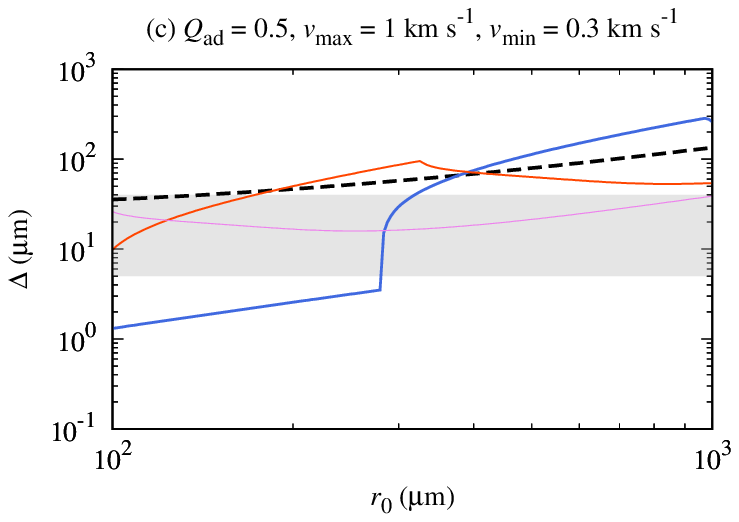}
\includegraphics[width=\columnwidth]{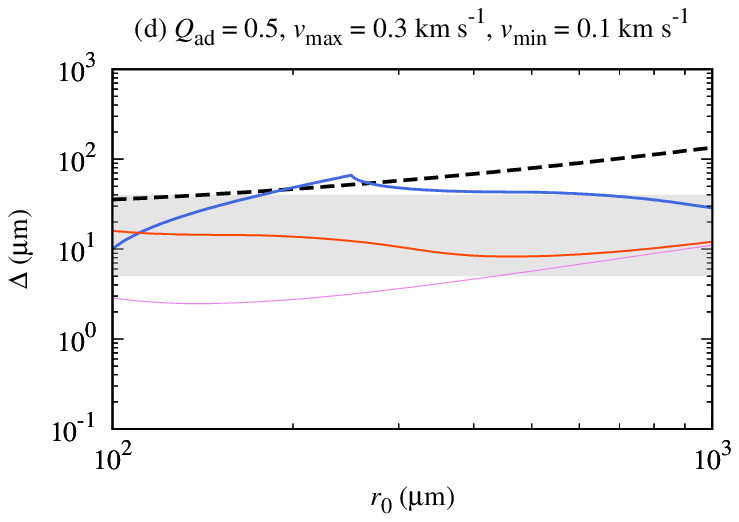}
\caption{
Thickness of FGRs, $\Delta$, as a function of chondrule radius, $r_{0}$.
Fine-grained rim formation without erosion is assumed: $Q_{\rm er} = 0$.
The black dashed line indicates the relationship between $\Delta$ and $r_{\rm 0}$ for chondrules in Murchison CM chondrite: ${\left( \Delta / 1\ \si{\micro}{\rm m} \right)} = 0.11 {\left( r_{0} / 1\ \si{\micro}{\rm m} \right)} + 24.5$ \citep{2018E&PSL.481..201H}. 
The gray shaded range indicates the typical thickness of FGRs around chondrules in unequilibrated ordinary chondrites: $5\ \si{\micro}{\rm m} \le \Delta \le 40\ \si{\micro}{\rm m}$ \citep{1984PolRe..35..126M}.
}
\label{fig:Delta-r-no-erosion}
\end{figure*}

\begin{figure*}
\centering
\includegraphics[width=\columnwidth]{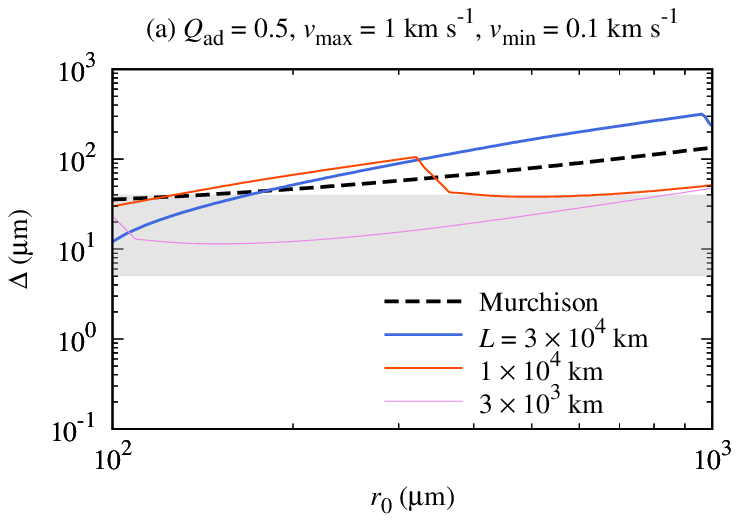}
\includegraphics[width=\columnwidth]{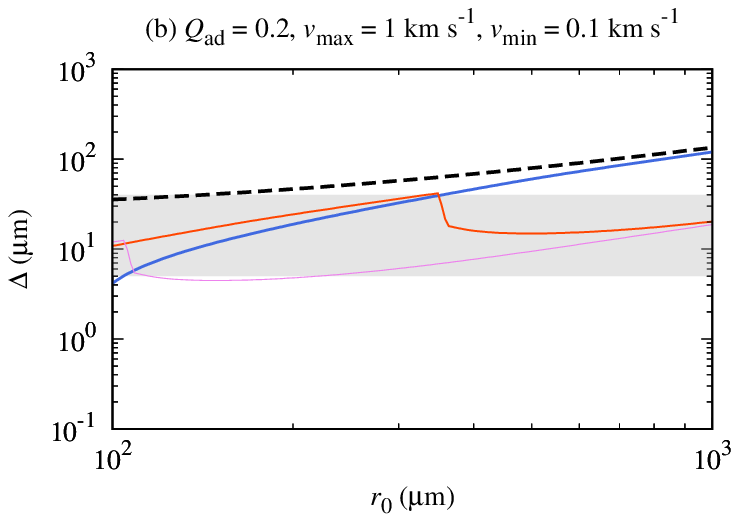}
\includegraphics[width=\columnwidth]{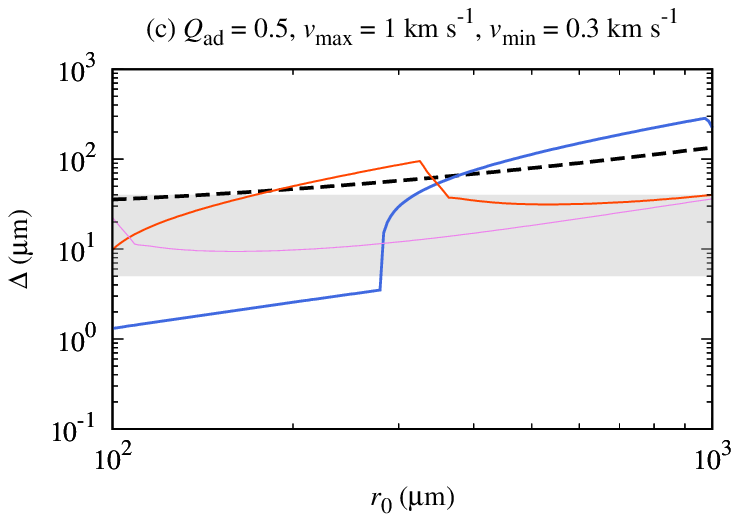}
\includegraphics[width=\columnwidth]{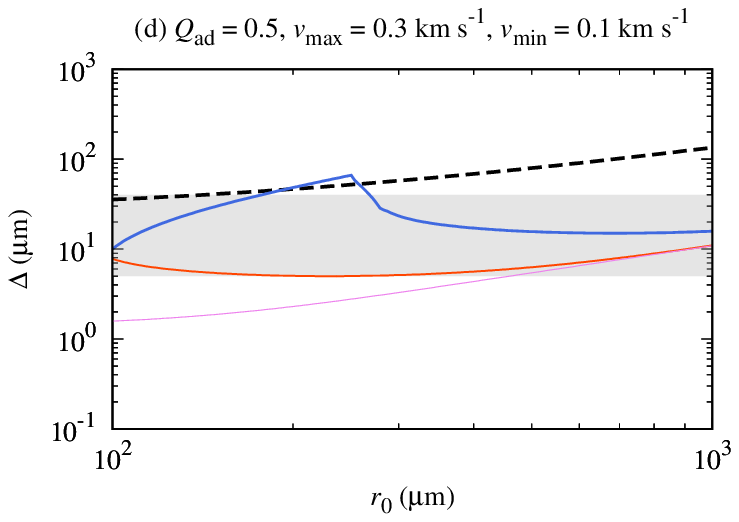}
\caption{
Thickness of FGRs, $\Delta$, as a function of chondrule radius, $r_{0}$.
Fine-grained rim formation with erosion is assumed: $Q_{\rm er} = -1$.
The black dashed line indicates the relationship between $\Delta$ and $r_{\rm 0}$ for chondrules in Murchison CM chondrite: ${\left( \Delta / 1\ \si{\micro}{\rm m} \right)} = 0.11 {\left( r_{0} / 1\ \si{\micro}{\rm m} \right)} + 24.5$ \citep{2018E&PSL.481..201H}. 
The gray shaded range indicates the typical thickness of FGRs around chondrules in unequilibrated ordinary chondrites: $5\ \si{\micro}{\rm m} \le \Delta \le 40\ \si{\micro}{\rm m}$ \citep{1984PolRe..35..126M}.
}
\label{fig:Delta-r-erosion}
\end{figure*}

Finally, we show the dependence of final rim thickness on chondrule radius.
Figure \ref{fig:Delta-r-no-erosion} shows the results for the case of $Q_{\rm er} = 0$ (rim formation without erosion) and Figure \ref{fig:Delta-r-erosion} is for the case of $Q_{\rm er} = -1$ (rim formation with erosion).
As shown in Figures \ref{fig:Delta-x-no-erosion} and \ref{fig:Delta-x-erosion}, FGR formation finishes at $x \sim 10^{5}\ {\rm km}$ because $v_{\rm imp} < v_{\rm min}$ for $x \gg 10^{5}\ {\rm km}$.
Then we stop numerical simulations at $x = 10^{6}\ {\rm km}$ in this study.

Figure \ref{fig:Delta-r-no-erosion}(a) shows the results for the case of $Q_{\rm ad} = 0.5$, $Q_{\rm er} = 0$, $v_{\rm max} = 1\ {\rm km}\ {\rm s}^{-1}$, and $v_{\rm min} = 0.1\ {\rm km}\ {\rm s}^{-1}$.
We found that the final rim thickness is approximately consistent with that for chondrules in Murchison CM chondrite: ${\left( \Delta / 1\ \si{\micro}{\rm m} \right)} = 0.11 {\left( r_{0} / 1\ \si{\micro}{\rm m} \right)} + 24.5$ \citep{2018E&PSL.481..201H}.
The value of $\Delta$ also depends on the spatial scale of the shock, $L$, and our numerical results show a good agreement with observations for CM chondrites when $L = 1 \times 10^{4}\ {\rm km}$ or $3 \times 10^{4}\ {\rm km}$.

Figure \ref{fig:Delta-r-no-erosion}(b) shows the results for the case of $Q_{\rm ad} = 0.2$, $Q_{\rm er} = 0$, $v_{\rm max} = 1\ {\rm km}\ {\rm s}^{-1}$, and $v_{\rm min} = 0.1\ {\rm km}\ {\rm s}^{-1}$.
As the accretion rate of FGRs is proportional to $Q_{\rm ad}$, the final thickness of FGRs formed in this setting is smaller than that shown in Figure \ref{fig:Delta-r-no-erosion}(a).
We found that the final rim thickness is in the range of $5\ \si{\micro}{\rm m} \le \Delta \le 40\ \si{\micro}{\rm m}$ for the cases of $L = 1 \times 10^{4}\ {\rm km}$ and $3 \times 10^{3}\ {\rm km}$.
This is consistent with the thickness of FGRs around chondrules in unequilibrated ordinary chondrites \citep{1984PolRe..35..126M}.
The observations by \citet{1984PolRe..35..126M} indicate that the thickness of FGRs is not dependent on the chondrule radius, and similar results are also reported by \citet{bigolski2017formation}.

We note that our results are based on simple one-dimensional simulations.
However, in reality, shock waves caused by eccentric planetesimals are bow shocks.
The trajectories of chondrules are curved and strongly depend on their size \citep[e.g.,][]{2013ApJ...776..101B,katsuda}.
Moreover, we assumed that the coefficient for adhesion is constant in the range of $v_{\rm min} < v_{\rm imp} < v_{\rm max}$; this assumption also seems to be unlikely.
For these reasons, we do not discuss the detailed features of the dependence of $\Delta$ on $r_{0}$ in this study.

Figure \ref{fig:Delta-r-no-erosion}(c) shows the results for the case of $Q_{\rm ad} = 0.5$, $Q_{\rm er} = 0$, $v_{\rm max} = 1\ {\rm km}\ {\rm s}^{-1}$, and $v_{\rm min} = 0.3\ {\rm km}\ {\rm s}^{-1}$.
Interestingly, the thickness of FGRs is significantly smaller than the observed values when $L = 3 \times 10^{4}\ {\rm km}$ and $r_{0} < 300\ \si{\micro}{\rm m}$.
This is because the maximum value of $v_{\rm imp}$ in Stage 2 is lower than $0.3\ {\rm km}\ {\rm s}^{-1}$ if the radius of chondrules is smaller than $300\ \si{\micro}{\rm m}$, as shown in Figure \ref{fig:vimp}(a).
In this case, FGRs cannot be formed in Stage 2 and final thickness would be equal to the thickness formed in Stage 1.

Figure \ref{fig:Delta-r-no-erosion}(d) shows the results for the case of $Q_{\rm ad} = 0.5$, $Q_{\rm er} = 0$, $v_{\rm max} = 0.3\ {\rm km}\ {\rm s}^{-1}$, and $v_{\rm min} = 0.1\ {\rm km}\ {\rm s}^{-1}$.
Although the final thickness of FGRs is smaller than that formed in Figure \ref{fig:Delta-r-no-erosion}(a), FGRs with thickness of 10--100 \si{\micro}m are formed even if $v_{\rm max} = 0.3\ {\rm km}\ {\rm s}^{-1}$.
In conclusion, the kinetic dust aggregation in shock waves around evaporating icy planetesimals would be the leading candidate for the origin of FGRs around chondrules in primitive chondrites.

Figure \ref{fig:Delta-r-erosion} shows the results for the case of FGR formation with erosion ($Q_{\rm er} = -1$).
Although the final thickness of FGRs formed in Figure \ref{fig:Delta-r-erosion} is slightly smaller than that in Figure \ref{fig:Delta-r-no-erosion} ($Q_{\rm er} = 0$), the general trends are similar and FGRs with thickness of 10--100 \si{\micro}m are formed even if we consider the effect of erosion.
This is consistent with the fact that the thickness of FGRs formed in Stage 1 is significantly smaller than that formed in Stage 2.

The relation between the thickness of FGRs and the radius of chondrules is discussed so far.
For chondrules in carbonaceous chondrites, the positive correlation was reported within the range of $100\ \si{\micro}{\rm m} < r_{0} < 1000\ \si{\micro}{\rm m}$ \citep[e.g.,][]{2018E&PSL.481..201H}.
In contrast, no clear correlation between $\Delta$ and $r_{0}$ was found for chondrules in unequilibrated ordinary chondrites \citep{1984PolRe..35..126M}.
Our results show that the positive correlation appears when accretion of FGRs occurs in the almost all region of Stage 2 (see Figure \ref{fig:Delta-x-no-erosion}(a)).

\section{discussion}

\subsection{Rim thickness formed in Stage 1: deceleration region behind the shock front}

As mentioned above, the thickness of FGRs formed in Stage 1 is significantly smaller than that formed in Stage 2.
Here we derive an analytic solution for the thickness of FGRs formed in Stage 1.
The motion of chondrules in Stage 1 is described as the deceleration behind the shock front.

Here we consider the accretion of fine dust grains onto chondrules in Stage 1, and we assume that $v_{\rm g}$, $\rho_{\rm g}$, and $c_{\rm s}$ are almost constant for simplicity.
Although the relative velocity of chondrules with respect to gas is supersonic at $x \ll l_{\rm stop}$, FGRs are not formed in this region because $v_{\rm imp}$ is higher than the maximum velocity for adhesion, $v_{\rm max}$.
Then $v_{\rm imp}$ will drop to the range for adhesion, and FGR formation in Stage 1 will start.

When the relative velocity of chondrules with respect to gas is subsonic, the time evolution of $v_{\rm imp}$ is given by
\begin{eqnarray}
\frac{{\rm d}v_{\rm imp}}{{\rm d}t} & \simeq & - {\left| \frac{{\rm d}v}{{\rm d}t} \right|} \nonumber \\
                                    & \simeq & - \frac{1}{0.64} \frac{\rho_{\rm g}}{\rho} \frac{c_{\rm s} v_{\rm imp}}{r_{0}}.
\end{eqnarray}
For the case of $v_{\rm min} < v_{\rm imp} < v_{\rm max}$, the time evolution of the radius of rimmed chondrules is given by
\begin{eqnarray}
\frac{{\rm d}r}{{\rm d}t} & = & \frac{Q_{\rm ad}}{4} \frac{\rho_{\rm d}}{\rho} v_{\rm imp} \nonumber \\
                          & \simeq & - \frac{0.64 Q_{\rm ad}}{4} \chi \frac{r_{0}}{c_{\rm s}} \frac{{\rm d}v_{\rm imp}}{{\rm d}t}.
\end{eqnarray}
Then the thickness of FGRs formed in Stage 1 would be approximately given by the following equation:
{\footnotesize
\begin{eqnarray}
\Delta_{1} & = & \frac{0.64 Q_{\rm ad}}{4} \chi \frac{v_{\rm max} - v_{\rm min}}{c_{\rm s}} r_{0} \nonumber \\
                      & \simeq & 2\ {\left( \frac{Q_{\rm ad}}{0.5} \right)} {\left( \frac{\chi}{1} \right)} {\left( \frac{v_{\rm max} - v_{\rm min}}{900\ {\rm m}\ {\rm s}^{-1}} \right)} {\left( \frac{r_{0}}{100\ \si{\micro}{\rm m}} \right)}\ \si{\micro}{\rm m}.
\end{eqnarray}
}

Our analytic solution suggests that the thickness of FGRs formed in stage 1 is $\Delta_{1} \simeq 2\ {\left( {r_{0}}/{100\ \si{\micro}{\rm m}} \right)}\ \si{\micro}{\rm m}$, and this value is one order of magnitude smaller than the observed thickness of FGRs around chondrules in CM chondrites \citep[e.g.,][]{2018E&PSL.481..201H}.
Thus we need to consider the FGR formation in Stage 2.

\subsection{Rim thickness formed in Stage 2: quasi-steady state in recovery region}

Similarly, we can derive the analytic solution for the thickness of FGRs formed in Stage 2.
When the spatial scale of the shock is sufficiently larger than the stopping length ($L \gg l_{\rm stop}$), the motion of chondrules in Stage 2 is described as the dynamically quasi-steady state.
In this region, the velocities of both gas and chondrules recover (see Equation \ref{eq:vg}), and the relative velocity of the chondrule to the gas is negligibly smaller than $v_{\rm g}$ \citep[see also][]{2019ApJ...877...84A}.

When we consider the quasi-steady state for the dynamics of chondrules in Stage 2, the differential of the velocity of chondrules is approximately given by the following equation:
\begin{eqnarray}
{\left| \frac{{\rm d}v}{{\rm d}x} \right|} & = & \frac{v}{l_{\rm stop}} \nonumber \\
                                           & \simeq & \frac{v_{\rm g}}{l_{\rm stop}} \nonumber \\
                                           & \simeq & \frac{1}{0.64} \frac{\rho_{\rm g}}{\rho} \frac{c_{\rm s}}{v_{\rm g}} \frac{v_{\rm imp}}{r_{0}}.
\end{eqnarray}
On the other hand, the differential of the velocity of gas is given as follows (see Equation \ref{eq:vg}):
\begin{equation}
{\left| \frac{{\rm d}v_{\rm g}}{{\rm d}x} \right|} = \frac{\left| v_{\rm g} - v_{0} \right|}{L}.
\end{equation}
Assuming that ${{\rm d}v} / {{\rm d}x}$ and ${{\rm d}v_{\rm g}} / {{\rm d}x}$ are approximately equal, the relative velocity of the chondrule from the gas, which is equal to $v_{\rm imp}$, is derived as follows:
\begin{equation}
v_{\rm imp} \simeq 0.64 \frac{\rho}{\rho_{\rm g}} \frac{v_{\rm g}}{c_{\rm s}} \frac{\left| v_{\rm g} - v_{0} \right|}{L} r_{0}.
\end{equation}
As $v_{\rm imp}$ takes the maximum at around $x \sim L$, we show the value of $v_{\rm imp}$ at $x = L$ as a reference: 
{\footnotesize
\begin{eqnarray}
v_{\rm imp}|_{x = L} \simeq & & 120\ {\left( \frac{\rho_{{\rm g}, 0}}{5 \times 10^{-10}\ {\rm g}\ {\rm cm}^{-3}} \right)}^{-1} \nonumber \\
                            & & \times {\left( \frac{L}{3 \times 10^{4}\ {\rm km}} \right)}^{-1} {\left( \frac{r_{0}}{100\ \si{\micro}{\rm m}} \right)}\ {\rm m}\ {\rm s}^{-1}. 
\end{eqnarray}
}

Then we can calculate the time evolution of the radius of rimmed chondrules.
When the impact velocity of fine dust grains satisfies $v_{\rm min} < v_{\rm imp} < v_{\rm max}$, the differential of the radius of rimmed chondrules is given by
\begin{eqnarray}
\frac{{\rm d}r}{{\rm d}x} & = & \frac{Q_{\rm ad}}{4} \frac{\rho_{\rm d}}{\rho} \frac{v_{\rm imp}}{v} \nonumber \\
                          & \simeq & \frac{0.64 Q_{\rm ad}}{4} \chi \frac{\left| v_{\rm g} - v_{0} \right|}{c_{\rm s}} \frac{r_{0}}{L}.
\end{eqnarray}
The maximum thickness formed in Stage 2, $\Delta_{2, {\rm max}}$, is therefore given by the following equation:
\begin{eqnarray}
\Delta_{2, {\rm max}} & = & \int_{0}^{\infty}\ {\rm d}x\ \frac{{\rm d}r}{{\rm d}x} \nonumber \\
                      & \simeq & 32\ {\left( \frac{Q_{\rm ad}}{0.5} \right)} {\left( \frac{\chi}{1} \right)} {\left( \frac{r_{0}}{100\ \si{\micro}{\rm m}} \right)}\ \si{\micro}{\rm m}.
\end{eqnarray}
We found that $\Delta_{2, {\rm max}} \gg \Delta_{1}$, thus FGRs would be mainly formed in Stage 2, quasi-steady state in recovery region.
The maximum thickness of FGRs formed in stage 2 is $\Delta_{2, {\rm max}} \simeq 32\ {\left( {r_{0}}/{100\ \si{\micro}{\rm m}} \right)}\ \si{\micro}{\rm m}$, and this value can explain the existence of thick FGRs around chondrules found in CM chondrites \citep[e.g.,][]{2018E&PSL.481..201H}.

We note that the thickness of FGRs formed in Stage 2 is approximately equal to $\Delta_{2, {\rm max}}$ only when $v_{\rm min} < v_{\rm imp}|_{x = L} < v_{\rm max}$.
When $v_{\rm imp}|_{x = L} \gg v_{\rm max}$, the thickness of FGRs is smaller than $\Delta_{2, {\rm max}}$ because fine dust grains cannot accrete onto chondrules at around $x \sim L$.
This effect appears in the blue line in Figures \ref{fig:Delta-r-no-erosion}(d) and \ref{fig:Delta-r-erosion}(d); FGRs around chondrules with radius larger than $0.25\ {\rm mm}$ are thinner than $\Delta_{2, {\rm max}}$.
In addition, FGRs are not formed in Stage 2 when $v_{\rm imp}|_{x = L} \ll v_{\rm min}$.

We also note that the power-law exponent for the relation between $\Delta$ and $r_{0}$ (for chondrules in carbonaceous chondrites) is still under debate.
Although several studies \citep[e.g.,][]{1992GeCoA..56.2873M,2004Icar..168..484C} reported that $\Delta$ is approximately proportional to $r_{0}$, \citet{2018E&PSL.481..201H} pointed out that $\Delta$ is approximately proportional to the square root of $r_{0}$.
When accretion of FGRs occurs in the entire region of Stage 2, our model predicts that $\Delta$ is proportional to ${r_{0}}^{1 - \beta}$, where $\beta$ is the exponent for the velocity dependence of $Q_{\rm ad}$ (i.e., $Q_{\rm ad}$ is proportional to ${v_{\rm imp}}^{- \beta}$).
Thus the relation between $\Delta$ and $r_{0}$ could be reproduced if $\beta \simeq 0.5$ in the range of $v_{\rm min} < v_{\rm imp} < v_{\rm max}$.
Although we set $\beta = 0$ (i.e., $Q_{\rm ad}$ is constant) in this preliminary study, we need to investigate the velocity dependence of $Q_{\rm ad}$ from laboratory experiments.

\subsection{Co-existence of rimmed and unrimmed chondrules}

Although FGRs are frequently observed around chondrules in primitive chondrites, the occurrence rate is not 100\%.
For unequilibrated ordinary chondrites, the occurrence rate is 79\% for Semarkona, 70\% for Watonga, and 59\% for Bishunpur \citep{2017LPICo1987.6234B}.
In addition, the occurrence rate of FGRs is only 15--20\% for Allende CV chondrite \citep{2018E&PSL.494...69S}.
Therefore, we must give an explanation for the co-existence of rimmed and unrimmed chondrules in the context of FGR formation.
Several mechanisms were proposed so far: \citet{2010GeCoA..74.4438T} claimed that unrimmed chondrules have lost FGRs during the brecciation process on their parent bodies, whereas \citet{2021A&A...652A..40U} proposed that unrimmed chondrules were formed via collisional fragmentation of chondritic aggregates in the solar nebula.

In our scenario, FGRs are formed via the kinetic dust aggregation process in the dusty region formed behind the evaporating icy planetesimal.
We note that dusty regions would be formed only when shock waves are caused by undifferentiated icy planetesimals; no dusty regions are expected for the case of differentiated planetesimals.
Therefore, if chondrules are formed via shock-wave heating events caused by both undifferentiated and differentiated planetesimals, we can expect the co-existence of rimmed and unrimmed chondrules.
As the critical diameter of icy planetesimals for differentiation would be approximately 1000 km, parts of chondrules might be formed via shock waves caused by huge planetesimals (or protoplanets) whose diameter is far larger than 1000 km.

\subsection{The oxygen isotope ratios and Mg\# systematics of chondrules}

The Mg\# of chondrules, which is defined as Mg\# = [MgO] / [MgO + FeO] in molar percent, reflects the oxidation state of iron during chondrule formation, and we can estimate the environment of chondrule formation (e.g., oxygen fugacity) from the Mg\#.
The mass-independent oxygen isotope fractionation, ${\Delta}^{17}$O, is also useful to estimate the redox conditions and dust-to-ice mass ratio in chondrule formation environment \citep[e.g.,][]{2015GeCoA.148..228T,2018GeCoA.224..116H,2020PNAS..11723426W}.
\citet{2015GeCoA.148..228T} calculated the dust-to-gas and dust-to-ice mass ratios in chondrule formation environment for chondrules in CR chondrites.
Using the mass balance and the equilibrium condensation model, they reported that type I (Mg\# $>$ 90) chondrules would be formed in moderately dust-rich environments (100--200 times the solar metallicity) and from ice-dust mixtures with 0--0.8 times the abundance of ice in CI chondrites.
Similar results are also reported by \citet{2018GeCoA.224..116H} for type I chondrules in CV chondrites.

When chondrules formed via bow shocks around evaporating undifferentiated icy planetesimals, Equation (\ref{eq:chi}) predicted that the degree of dust enrichment would be on the order of 100 (i.e., the dust-to-gas mass ratio is on the order of 1).
This value is approximately consistent with the results from Mg\#--${\Delta}^{17}$O systematics for type I chondrules in carbonaceous chondrites \citep[e.g.,][]{2020PNAS..11723426W}.
The dust-to-ice mass ratio in chondrule formation environment would be approximately equal to the bulk composition of the planetesimals.
Therefore, undifferentiated icy planetesimals with slightly dust-rich compared to the CI composition might be suitable to reproduce the oxygen isotope ratios and Mg\# systematics.
We will discuss the redox conditions and dust-to-ice mass ratio in chondrule formation environment in future studies.

\section{summary}

FGRs are frequently found around chondrules in primitive chondrites.
The remarkable feature of FGRs is their submicron-sized and non-porous nature \citep[e.g.,][]{2006GeCoA..70.1271T,2008GeCoA..72..602C}.
The typical thickness of FGRs around chondrules is 10--100 \si{\micro}m.

\editX{Recently,} \citet{2019GeCoA.264..118L} proposed an idea for the origin of FGRs: high-speed collisions between chondrules and fine dust grains, which is called the kinetic dust aggregation process\editX{.
Experimental and numerical studies revealed that (sub)micron-sized ceramic particles can stick to a ceramic substrate in a vacuum, and the impact velocity for sticking is approximately 0.1--1 km/s} \citep[see][and references therein]{hanft2015overview}.
The resulting dust layer formed via the kinetic dust aggregation \editR{would} have low porosity and \editX{are} \editR{be} fine grained\editX{, as illustrated in Figure \ref{fig:Liffman}}.
Therefore, \editX{we can} \editR{it would be possible to} reproduce the observed structure of FGRs if they are formed via the kinetic dust aggregation process, which should be related to chondrule-forming supersonic events.

\editR{In this study,} \editX{We} \editR{we} examined the possibility of FGR formation via kinetic dust aggregation in chondrule-forming shock waves (see Figure \ref{fig:schematic}).
When shock waves are caused by undifferentiated icy planetesimals, fine dust grains would be released from the planetary surface due to evaporation of icy planetesimals \citep[e.g.,][]{2013ApJ...764..120T}.
\editX{The enrichment of fine dust grains in chondrule-forming environment would be preferred from a variety of perspectives.}
\editR{Then the dusty region would be formed behind the shock front.}
We studied the dynamics of chondrules behind the shock front \editR{using simple one-dimensional calculations}, and the growth of FGRs via kinetic dust aggregation was investigated.
Our key findings are summarized as follows.

\begin{enumerate}
\item{
\editR{As} \citet{2019ApJ...877...84A} \editX{found that} \editR{pointed out,} the dynamical evolution of chondrules in shock waves can be divided into two stages: deceleration region behind the shock front (Stage 1) and recovery region where the velocity of chondrules and gas approach the pre-shock velocity (Stage 2).
\editX{For the case of shock waves,} \editR{We showed that} $v_{\rm imp}$ is approximately proportional to $r_{0}$ in Stage 2.
\editX{We derived an analytical solution for $v_{\rm imp} = v_{\rm imp} {\left( r_{0} \right)}$ in Stage 2.}
}
\item{
We found that non-porous FGRs with the thickness of 10--100 \si{\micro}m are formed in shock waves around evaporating icy planetesimals (Figures \ref{fig:Delta-x-no-erosion} and \ref{fig:Delta-x-erosion}).
This thickness is in good agreement with observations \citep[e.g.,][]{1984PolRe..35..126M,2018E&PSL.481..201H}.
We also found that the thickness of FGRs formed in Stage 1 is significantly smaller than that formed in Stage 2.
}
\item{
We derived analytic solutions for the thickness of FGRs formed in Stages 1 and 2.
The motion of chondrules in Stage 1 is described as the deceleration behind the shock front, while the motion of chondrules in Stage 2 is described as the dynamically quasi-steady state.
Our analytical solutions also predict that the thickness of FGRs is proportional to the chondrule radius when the effect of erosion is negligible.
}
\item{
In some cases, the erosion of FGRs occurs but FGRs partly survive after erosion, and fine dust grains accrete onto chondrules again (see Figure \ref{fig:Delta-x-erosion}).
Thus multi-layered FGRs would be formed by single shock-heating event; this might be consistent with the fact that chondrules in some CM2 chondrites are covered by multi-layered FGRs \citep{1992GeCoA..56.2873M}.
}
\item{
Although FGRs are frequently observed around chondrules in primitive chondrites, the occurrence rate is not 100\%.
\editX{Therefore, we should give an explanation for the co-existence of rimmed and unrimmed chondrules.}
In our scenario, \editX{FGRs are formed via the kinetic dust aggregation process} \editR{FGR formation would proceed} in the dusty region formed behind the evaporating icy planetesimal.
We note that dusty regions would be formed only when shock waves are caused by undifferentiated icy planetesimals; no dusty regions are expected for the case of differentiated planetesimals.
Therefore, if chondrules are formed via shock-wave heating events caused by both undifferentiated and differentiated planetesimals, we can expect the co-existence of rimmed and unrimmed chondrules.
}
\end{enumerate}

\section*{acknowledgments}

\editR{The anonymous reviewer provided a constructive review that improved this paper.}
\editX{We} \editR{The authors} thank Yuji Matsumoto for helpful comments.
S.A.\ was supported by JSPS KAKENHI Grant No.\ JP20J00598.
T.N.\ was supported by JSPS KAKENHI Grant No.\ JP18K03721.


\bibliography{sample}

\begin{thebibliography}{}
\expandafter\ifx\csname natexlab\endcsname\relax\def\natexlab#1{#1}\fi
\providecommand{\url}[1]{\href{#1}{#1}}
\providecommand{\dodoi}[1]{doi:~\href{http://doi.org/#1}{\nolinkurl{#1}}}
\providecommand{\doeprint}[1]{\href{http://ascl.net/#1}{\nolinkurl{http://ascl.net/#1}}}
\providecommand{\doarXiv}[1]{\href{https://arxiv.org/abs/#1}{\nolinkurl{https://arxiv.org/abs/#1}}}

\bibitem[{{Akedo}(2006)}]{akedo2006aerosol}
{Akedo}, J. 2006, Journal of the American Ceramic Society, 89, 1834

\bibitem[{{Akedo}(2008)}]{akedo2008room}
---. 2008, Journal of Thermal Spray Technology, 17, 181

\bibitem[{{Akedo} {et~al.}(2008){Akedo}, {Nakano}, {Park}, {Baba}, \&
  {Ashida}}]{akedo2008aerosol}
{Akedo}, J., {Nakano}, S., {Park}, J., {Baba}, S., \& {Ashida}, K. 2008,
  Synthesiology English edition, 1, 121

\bibitem[{{Alexander} {et~al.}(2008){Alexander}, {Grossman}, {Ebel}, \&
  {Ciesla}}]{2008Sci...320.1617A}
{Alexander}, C.~M.~O.~D., {Grossman}, J.~N., {Ebel}, D.~S., \& {Ciesla}, F.~J.
  2008, Science, 320, 1617, \dodoi{10.1126/science.1156561}

\bibitem[{{Arakawa}(2017)}]{2017ApJ...846..118A}
{Arakawa}, S. 2017, The Astrophysical Journal, 846, 118,
  \dodoi{10.3847/1538-4357/aa8564}

\bibitem[{{Arakawa} \& {Nakamoto}(2019)}]{2019ApJ...877...84A}
{Arakawa}, S., \& {Nakamoto}, T. 2019, The Astrophysical Journal, 877, 84,
  \dodoi{10.3847/1538-4357/ab1b3e}

\bibitem[{{Arakawa} \& {Ohno}(2020)}]{2020MNRAS.497.1166A}
{Arakawa}, S., \& {Ohno}, K. 2020, Monthly Notices of the Royal Astronomical
  Society, 497, 1166, \dodoi{10.1093/mnras/staa2031}

\bibitem[{{Beitz} {et~al.}(2013{\natexlab{a}}){Beitz}, {Blum}, {Mathieu},
  {Pack}, \& {Hezel}}]{2013GeCoA.116...41B}
{Beitz}, E., {Blum}, J., {Mathieu}, R., {Pack}, A., \& {Hezel}, D.~C.
  2013{\natexlab{a}}, Geochimica et Cosmochimica Acta, 116, 41,
  \dodoi{10.1016/j.gca.2012.04.059}

\bibitem[{{Beitz} {et~al.}(2013{\natexlab{b}}){Beitz}, {G{\"u}ttler},
  {Nakamura}, {Tsuchiyama}, \& {Blum}}]{2013Icar..225..558B}
{Beitz}, E., {G{\"u}ttler}, C., {Nakamura}, A.~M., {Tsuchiyama}, A., \& {Blum},
  J. 2013{\natexlab{b}}, Icarus, 225, 558, \dodoi{10.1016/j.icarus.2013.04.028}

\bibitem[{Bigolski(2017)}]{bigolski2017formation}
Bigolski, J.~N. 2017, The Formation of Fine-Grained Chondrule Rims in
  Unequilibrated Ordinary Chondrites (City University of New York (Ph.D.
  thesis))

\bibitem[{{Bigolski} \& {Weisberg}(2017)}]{2017LPICo1987.6234B}
{Bigolski}, J.~N., \& {Weisberg}, M.~K. 2017, in 80th Annual Meeting of the
  Meteoritical Society, Vol.~80, 6234

\bibitem[{{Birnstiel} {et~al.}(2018){Birnstiel}, {Dullemond}, {Zhu}, {Andrews},
  {Bai}, {Wilner}, {Carpenter}, {Huang}, {Isella}, {Benisty}, {P{\'e}rez}, \&
  {Zhang}}]{2018ApJ...869L..45B}
{Birnstiel}, T., {Dullemond}, C.~P., {Zhu}, Z., {et~al.} 2018, The
  Astrophysical Journal Letters, 869, L45, \dodoi{10.3847/2041-8213/aaf743}

\bibitem[{{Bland} {et~al.}(2011){Bland}, {Howard}, {Prior}, {Wheeler}, {Hough},
  \& {Dyl}}]{2011NatGe...4..244B}
{Bland}, P.~A., {Howard}, L.~E., {Prior}, D.~J., {et~al.} 2011, Nature
  Geoscience, 4, 244, \dodoi{10.1038/ngeo1120}

\bibitem[{{Boley} {et~al.}(2013){Boley}, {Morris}, \&
  {Desch}}]{2013ApJ...776..101B}
{Boley}, A.~C., {Morris}, M.~A., \& {Desch}, S.~J. 2013, The Astrophysical
  Journal, 776, 101, \dodoi{10.1088/0004-637X/776/2/101}

\bibitem[{{Brown}(2012)}]{2012AREPS..40..467B}
{Brown}, M.~E. 2012, Annual Review of Earth and Planetary Sciences, 40, 467,
  \dodoi{10.1146/annurev-earth-042711-105352}

\bibitem[{{Chizmadia} \& {Brearley}(2008)}]{2008GeCoA..72..602C}
{Chizmadia}, L.~J., \& {Brearley}, A.~J. 2008, Geochimica et Cosmochimica Acta,
  72, 602, \dodoi{10.1016/j.gca.2007.10.019}

\bibitem[{{Ciesla} {et~al.}(2004){Ciesla}, {Hood}, \&
  {Weidenschilling}}]{2004M&PS...39.1809C}
{Ciesla}, F.~J., {Hood}, L.~L., \& {Weidenschilling}, S.~J. 2004, Meteoritics
  \& Planetary Science, 39, 1809, \dodoi{10.1111/j.1945-5100.2004.tb00077.x}

\bibitem[{{Cuzzi}(2004)}]{2004Icar..168..484C}
{Cuzzi}, J.~N. 2004, Icarus, 168, 484, \dodoi{10.1016/j.icarus.2003.12.008}

\bibitem[{{D'Alessio} {et~al.}(2001){D'Alessio}, {Calvet}, \&
  {Hartmann}}]{2001ApJ...553..321D}
{D'Alessio}, P., {Calvet}, N., \& {Hartmann}, L. 2001, The Astrophysical
  Journal, 553, 321, \dodoi{10.1086/320655}

\bibitem[{{Daneshian} \& {Assadi}(2014)}]{2014JTST...23..541D}
{Daneshian}, B., \& {Assadi}, H. 2014, Journal of Thermal Spray Technology, 23,
  541, \dodoi{10.1007/s11666-013-0019-4}

\bibitem[{{Fulle} {et~al.}(2019){Fulle}, {Blum}, {Green}, {Gundlach},
  {Herique}, {Moreno}, {Mottola}, {Rotundi}, \&
  {Snodgrass}}]{2019MNRAS.482.3326F}
{Fulle}, M., {Blum}, J., {Green}, S.~F., {et~al.} 2019, Monthly Notices of the
  Royal Astronomical Society, 482, 3326, \dodoi{10.1093/mnras/sty2926}

\bibitem[{{Grundy} {et~al.}(2019){Grundy}, {Noll}, {Buie}, {Benecchi},
  {Ragozzine}, \& {Roe}}]{2019Icar..334...30G}
{Grundy}, W.~M., {Noll}, K.~S., {Buie}, M.~W., {et~al.} 2019, Icarus, 334, 30,
  \dodoi{10.1016/j.icarus.2018.12.037}

\bibitem[{{Hanft} {et~al.}(2015){Hanft}, {Exner}, {Schubert}, {St{\"o}cker},
  {Fuierer}, \& {Moos}}]{hanft2015overview}
{Hanft}, D., {Exner}, J., {Schubert}, M., {et~al.} 2015, Journal of Ceramic
  Science and Technology, 6, 147

\bibitem[{{Hanna} \& {Ketcham}(2018)}]{2018E&PSL.481..201H}
{Hanna}, R.~D., \& {Ketcham}, R.~A. 2018, Earth and Planetary Science Letters,
  481, 201, \dodoi{10.1016/j.epsl.2017.10.029}

\bibitem[{{Hertwig} {et~al.}(2018){Hertwig}, {Defouilloy}, \&
  {Kita}}]{2018GeCoA.224..116H}
{Hertwig}, A.~T., {Defouilloy}, C., \& {Kita}, N.~T. 2018, Geochimica et
  Cosmochimica Acta, 224, 116, \dodoi{10.1016/j.gca.2017.12.013}

\bibitem[{{Hewins} {et~al.}(2012){Hewins}, {Zanda}, \&
  {Bendersky}}]{2012GeCoA..78....1H}
{Hewins}, R.~H., {Zanda}, B., \& {Bendersky}, C. 2012, Geochimica et
  Cosmochimica Acta, 78, 1, \dodoi{10.1016/j.gca.2011.11.027}

\bibitem[{{Hood} \& {Horanyi}(1991)}]{1991Icar...93..259H}
{Hood}, L.~L., \& {Horanyi}, M. 1991, Icarus, 93, 259,
  \dodoi{10.1016/0019-1035(91)90211-B}

\bibitem[{{Jacquet} \& {Thompson}(2014)}]{2014ApJ...797...30J}
{Jacquet}, E., \& {Thompson}, C. 2014, The Astrophysical Journal, 797, 30,
  \dodoi{10.1088/0004-637X/797/1/30}

\bibitem[{{Johnson} {et~al.}(2014){Johnson}, {Glaser}, {Cheng}, {Kub}, \&
  {Eddy}}]{2014APExp...7c5501J}
{Johnson}, S.~D., {Glaser}, E.~R., {Cheng}, S.-F., {Kub}, F.~J., \& {Eddy},
  Charles~R., J. 2014, Applied Physics Express, 7, 035501,
  \dodoi{10.7567/APEX.7.035501}

\bibitem[{{Kaneko} {et~al.}(2022){Kaneko}, {Arakawa}, \&
  {Nakamoto}}]{2022Icar..37414726K}
{Kaneko}, H., {Arakawa}, S., \& {Nakamoto}, T. 2022, Icarus, 374, 114726,
  \dodoi{10.1016/j.icarus.2021.114726}

\bibitem[{{Katsuda}(2017)}]{katsuda}
{Katsuda}, Y. 2017, {Planetesimal Bow Shocks with High Dust-to-Gas Mass Ratio:
  A Possible Chondrule Formation Site} (Tokyo Institute of Technology (master
  thesis))

\bibitem[{{Lauretta} {et~al.}(2000){Lauretta}, {Hua}, \&
  {Buseck}}]{2000GeCoA..64.3263L}
{Lauretta}, D.~S., {Hua}, X., \& {Buseck}, P.~R. 2000, Geochimica et
  Cosmochimica Acta, 64, 3263, \dodoi{10.1016/S0016-7037(00)00425-7}

\bibitem[{{Liffman}(2019)}]{2019GeCoA.264..118L}
{Liffman}, K. 2019, Geochimica et Cosmochimica Acta, 264, 118,
  \dodoi{10.1016/j.gca.2019.08.009}

\bibitem[{{Mai} {et~al.}(2018){Mai}, {Desch}, {Boley}, \&
  {Weiss}}]{2018ApJ...857...96M}
{Mai}, C., {Desch}, S.~J., {Boley}, A.~C., \& {Weiss}, B.~P. 2018, The
  Astrophysical Journal, 857, 96, \dodoi{10.3847/1538-4357/aab711}

\bibitem[{{Mann} {et~al.}(2016){Mann}, {Boley}, \&
  {Morris}}]{2016ApJ...818..103M}
{Mann}, C.~R., {Boley}, A.~C., \& {Morris}, M.~A. 2016, The Astrophysical
  Journal, 818, 103, \dodoi{10.3847/0004-637X/818/2/103}

\bibitem[{{Matsumoto} {et~al.}(2021){Matsumoto}, {Hasegawa}, {Matsuda}, \&
  {Liu}}]{2021Icar..36714538M}
{Matsumoto}, Y., {Hasegawa}, Y., {Matsuda}, N., \& {Liu}, M.-C. 2021, Icarus,
  367, 114538, \dodoi{10.1016/j.icarus.2021.114538}

\bibitem[{{Matsumoto} {et~al.}(2019){Matsumoto}, {Wakita}, {Hasegawa}, \&
  {Oshino}}]{2019ApJ...887..248M}
{Matsumoto}, Y., {Wakita}, S., {Hasegawa}, Y., \& {Oshino}, S. 2019, The
  Astrophysical Journal, 887, 248, \dodoi{10.3847/1538-4357/ab5b06}

\bibitem[{{Matsunami}(1984)}]{1984PolRe..35..126M}
{Matsunami}, S. 1984, National Institute Polar Research Memoirs, 35, 126

\bibitem[{{Metzler} {et~al.}(1992){Metzler}, {Bischoff}, \&
  {Stoeffler}}]{1992GeCoA..56.2873M}
{Metzler}, K., {Bischoff}, A., \& {Stoeffler}, D. 1992, Geochimica et
  Cosmochimica Acta, 56, 2873, \dodoi{10.1016/0016-7037(92)90365-P}

\bibitem[{{Morfill} {et~al.}(1998){Morfill}, {Durisen}, \&
  {Turner}}]{1998Icar..134..180M}
{Morfill}, G.~E., {Durisen}, R.~H., \& {Turner}, G.~W. 1998, Icarus, 134, 180,
  \dodoi{10.1006/icar.1998.5948}

\bibitem[{{Morris} {et~al.}(2012){Morris}, {Boley}, {Desch}, \&
  {Athanassiadou}}]{2012ApJ...752...27M}
{Morris}, M.~A., {Boley}, A.~C., {Desch}, S.~J., \& {Athanassiadou}, T. 2012,
  The Astrophysical Journal, 752, 27, \dodoi{10.1088/0004-637X/752/1/27}

\bibitem[{{Nagasawa} {et~al.}(2019){Nagasawa}, {Tanaka}, {Tanaka}, {Nomura},
  {Nakamoto}, \& {Miura}}]{2019ApJ...871..110N}
{Nagasawa}, M., {Tanaka}, K.~K., {Tanaka}, H., {et~al.} 2019, The Astrophysical
  Journal, 871, 110, \dodoi{10.3847/1538-4357/aaf795}

\bibitem[{{P{\"a}tzold} {et~al.}(2019){P{\"a}tzold}, {Andert}, {Hahn},
  {Barriot}, {Asmar}, {H{\"a}usler}, {Bird}, {Tellmann}, {Oschlisniok}, \&
  {Peter}}]{2019MNRAS.483.2337P}
{P{\"a}tzold}, M., {Andert}, T.~P., {Hahn}, M., {et~al.} 2019, Monthly Notices
  of the Royal Astronomical Society, 483, 2337, \dodoi{10.1093/mnras/sty3171}

\bibitem[{{Pollack} {et~al.}(1994){Pollack}, {Hollenbach}, {Beckwith},
  {Simonelli}, {Roush}, \& {Fong}}]{1994ApJ...421..615P}
{Pollack}, J.~B., {Hollenbach}, D., {Beckwith}, S., {et~al.} 1994, The
  Astrophysical Journal, 421, 615, \dodoi{10.1086/173677}

\bibitem[{{Schrader} {et~al.}(2013){Schrader}, {Connolly}, {Lauretta},
  {Nagashima}, {Huss}, {Davidson}, \& {Domanik}}]{2013GeCoA.101..302S}
{Schrader}, D.~L., {Connolly}, H.~C., {Lauretta}, D.~S., {et~al.} 2013,
  Geochimica et Cosmochimica Acta, 101, 302, \dodoi{10.1016/j.gca.2012.09.045}

\bibitem[{{Sears} {et~al.}(1993){Sears}, {Benoit}, \&
  {Jie}}]{1993Metic..28..669S}
{Sears}, D. W.~G., {Benoit}, P.~H., \& {Jie}, L. 1993, Meteoritics, 28, 669,
  \dodoi{10.1111/j.1945-5100.1993.tb00638.x}

\bibitem[{{Simon} {et~al.}(2018){Simon}, {Cuzzi}, {McCain}, {Cato},
  {Christoffersen}, {Fisher}, {Srinivasan}, {Tait}, {Olson}, \&
  {Scargle}}]{2018E&PSL.494...69S}
{Simon}, J.~I., {Cuzzi}, J.~N., {McCain}, K.~A., {et~al.} 2018, Earth and
  Planetary Science Letters, 494, 69, \dodoi{10.1016/j.epsl.2018.04.021}

\bibitem[{{Skorov} \& {Blum}(2012)}]{2012Icar..221....1S}
{Skorov}, Y., \& {Blum}, J. 2012, Icarus, 221, 1,
  \dodoi{10.1016/j.icarus.2012.01.012}

\bibitem[{{Takayama} \& {Tomeoka}(2012)}]{2012GeCoA..98....1T}
{Takayama}, A., \& {Tomeoka}, K. 2012, Geochimica et Cosmochimica Acta, 98, 1,
  \dodoi{10.1016/j.gca.2012.08.015}

\bibitem[{{Tanaka} {et~al.}(2013){Tanaka}, {Yamamoto}, {Tanaka}, {Miura},
  {Nagasawa}, \& {Nakamoto}}]{2013ApJ...764..120T}
{Tanaka}, K.~K., {Yamamoto}, T., {Tanaka}, H., {et~al.} 2013, The Astrophysical
  Journal, 764, 120, \dodoi{10.1088/0004-637X/764/2/120}

\bibitem[{{Tazaki} {et~al.}(2021){Tazaki}, {Murakawa}, {Muto}, {Honda}, \&
  {Inoue}}]{2021ApJ...910...26T}
{Tazaki}, R., {Murakawa}, K., {Muto}, T., {Honda}, M., \& {Inoue}, A.~K. 2021,
  The Astrophysical Journal, 910, 26, \dodoi{10.3847/1538-4357/abdd3d}

\bibitem[{{Tenner} {et~al.}(2015){Tenner}, {Nakashima}, {Ushikubo}, {Kita}, \&
  {Weisberg}}]{2015GeCoA.148..228T}
{Tenner}, T.~J., {Nakashima}, D., {Ushikubo}, T., {Kita}, N.~T., \& {Weisberg},
  M.~K. 2015, Geochimica et Cosmochimica Acta, 148, 228,
  \dodoi{10.1016/j.gca.2014.09.025}

\bibitem[{{Tomeoka} \& {Ohnishi}(2010)}]{2010GeCoA..74.4438T}
{Tomeoka}, K., \& {Ohnishi}, I. 2010, Geochimica et Cosmochimica Acta, 74,
  4438, \dodoi{10.1016/j.gca.2010.04.058}

\bibitem[{{Trigo-Rodriguez} {et~al.}(2006){Trigo-Rodriguez}, {Rubin}, \&
  {Wasson}}]{2006GeCoA..70.1271T}
{Trigo-Rodriguez}, J.~M., {Rubin}, A.~E., \& {Wasson}, J.~T. 2006, Geochimica
  et Cosmochimica Acta, 70, 1271, \dodoi{10.1016/j.gca.2005.11.009}

\bibitem[{{Umst{\"a}tter} \& {Urbassek}(2021)}]{2021A&A...652A..40U}
{Umst{\"a}tter}, P., \& {Urbassek}, H.~M. 2021, Astronomy \& Astrophysics, 652,
  A40, \dodoi{10.1051/0004-6361/202141581}

\bibitem[{{Visser} {et~al.}(2021){Visser}, {Dr{\c{a}}{\.z}kowska}, \&
  {Dominik}}]{2021A&A...647A.126V}
{Visser}, R.~G., {Dr{\c{a}}{\.z}kowska}, J., \& {Dominik}, C. 2021, Astronomy
  \& Astrophysics, 647, A126, \dodoi{10.1051/0004-6361/202039769}

\bibitem[{{Wahlberg Jansson} \& {Johansen}(2017)}]{2017MNRAS.469S.149W}
{Wahlberg Jansson}, K., \& {Johansen}, A. 2017, Monthly Notices of the Royal
  Astronomical Society, 469, S149, \dodoi{10.1093/mnras/stx1470}

\bibitem[{{Weidenschilling} {et~al.}(1998){Weidenschilling}, {Marzari}, \&
  {Hood}}]{1998Sci...279..681W}
{Weidenschilling}, S.~J., {Marzari}, F., \& {Hood}, L.~L. 1998, Science, 279,
  681, \dodoi{10.1126/science.279.5351.681}

\bibitem[{{Williams} {et~al.}(2020){Williams}, {Sanborn}, {Defouilloy}, {Yin},
  {Kita}, {Ebel}, {Yamakawa}, \& {Yamashita}}]{2020PNAS..11723426W}
{Williams}, C.~D., {Sanborn}, M.~E., {Defouilloy}, C., {et~al.} 2020,
  Proceedings of the National Academy of Science, 117, 23426,
  \dodoi{10.1073/pnas.2005235117}

\bibitem[{{Xiang} {et~al.}(2019){Xiang}, {Carballido}, {Hanna}, {Matthews}, \&
  {Hyde}}]{2019Icar..321...99X}
{Xiang}, C., {Carballido}, A., {Hanna}, R.~D., {Matthews}, L.~S., \& {Hyde},
  T.~W. 2019, Icarus, 321, 99, \dodoi{10.1016/j.icarus.2018.10.014}

\bibitem[{{Xiang} {et~al.}(2021){Xiang}, {Carballido}, {Matthews}, \&
  {Hyde}}]{2021Icar..35414053X}
{Xiang}, C., {Carballido}, A., {Matthews}, L.~S., \& {Hyde}, T.~W. 2021,
  Icarus, 354, 114053, \dodoi{10.1016/j.icarus.2020.114053}

\bibitem[{{Zanetta} {et~al.}(2021){Zanetta}, {Leroux}, {Le Guillou}, {Zanda},
  \& {Hewins}}]{2021GeCoA.295..135Z}
{Zanetta}, P.~M., {Leroux}, H., {Le Guillou}, C., {Zanda}, B., \& {Hewins},
  R.~H. 2021, Geochimica et Cosmochimica Acta, 295, 135,
  \dodoi{10.1016/j.gca.2020.12.015}

\end{thebibliography}



\end{document}